\documentclass[a4paper,useAMS,usenatbib,usegraphicx]{mn2e}

\usepackage{epsfig}

\input{psfig.sty}
\newcommand{\beq}{\begin{equation}}
\newcommand{\eeq}{\end{equation}}

\def\gtsima{$\; \buildrel > \over \sim \;$}
\def\ltsima{$\; \buildrel < \over \sim \;$}
\def\prosima{$\; \buildrel \propto \over \sim \;$}
\def\gsim{\lower.7ex\hbox{\gtsima}}
\def\lsim{\lower.7ex\hbox{\ltsima}}
\def\simgt{\lower.7ex\hbox{\gtsima}}
\def\simlt{\lower.7ex\hbox{\ltsima}}
\def\simpr{\lower.7ex\hbox{\prosima}}

\newcommand{\apj}{ApJ}
\newcommand{\apjl}{ApJL}
\newcommand{\apjs}{ApJS}
\newcommand{\aj}{AJ}
\newcommand{\mnras}{MNRAS}
\newcommand{\aap}{A\&A}

\newcommand{\pasp}{PASP}

\newdimen\hssize
\newdimen\hdsize
\usepackage{amssymb}
\usepackage{amsmath}

\begin{document}
\setlength{\hbadness}{10000}

\title[Mid-infrared emission in protostellar cores]
{Stochastic grain heating and mid-infrared emission in protostellar cores}

\author[Ya. N. Pavlyuchenkov et al.]
{Ya. N. Pavlyuchenkov$^{1}$\thanks{E-mail: pavyar@inasan.ru},
D. S. Wiebe$^{1}$, V. V. Akimkin$^{1}$, M. S. Khramtsova$^{1}$,
\newauthor
and Th. Henning$^{2}$ \\
$^{1}$Institute of Astronomy, Russian Academy of Sciences, Pyatnitskaya str. 48, Moscow 119017, Russia\\
$^{2}$Max Planck Institute for Astronomy, K\"onigstuhl 17, Heidelberg D69117, Germany}


\maketitle

\label{firstpage}

\begin{abstract}
Stochastic heating of small grains is often mentioned as a primary cause of
large infrared (IR) fluxes from star-forming galaxies, e.g. at 24\,$\mu$m. If
the mechanism does work at a galaxy-wide scale, it should show up at smaller
scales as well. We calculate temperature probability density distributions
within a model protostellar core for four  dust components: large silicate and
graphite grains, small graphite grains, and polycyclic aromatic hydrocarbon
particles. The corresponding spectral energy distributions are calculated and
compared with observations of a representative infrared dark cloud core. We show
that stochastic heating, induced by the standard interstellar radiation field,
cannot explain high mid-IR emission toward the centre of the core. In order to
reproduce the observed emission from the core projected centre, in particular,
at 24\,$\mu$m, we need to increase the ambient radiation field by a factor of
about 70. However, the model with enhanced radiation field predicts even higher
intensities at the core periphery, giving it a ring-like appearance, that is not
observed. We discuss possible implications of this finding and also discuss a
role of other non-radiative dust heating processes.
\end{abstract}


\section{Introduction}

Thanks to IRAS, ISO, and, in particular, the Spitzer Space Telescope,
mid-infrared emission now attracts significant attention both at small scale, as
an indicator of the protostar formation \citep[e.g., ][]{raganetal2009}, and at
large scale, as an indicator of the global star formation rate \citep[e.g.,
][]{calzetti2007,calzetti2010}. In both cases making reliable conclusions is
only possible with a radiation transfer (RT) model and a detailed account of the
dust thermal balance.

Numerous models have been developed to describe emergent spectra for
a dusty medium. Most of them treat dust radiative heating as a continuous
process that leads to a common equilibrium temperature for dust grains
\citep{csdust3,dusty,wolf,radmc,robitaille}.
However, it has been recognized long ago that grains of different sizes respond
differently to a photon absorption, depending on the ratio of the photon energy
and the grain thermal energy \citep{greenberg}. If the former is greater or
comparable to the latter, the temperature evolution of a single grain is
stochastic and consists of short temperature spikes upon a photon absorption,
followed by prolonged ``cold'' periods, when grain temperature is lower than the
equilibrium temperature it would have if heating were continuous
\citep{Duley1973}. Because of the energy relation, mentioned above, stochastic
heating depends on the grain size and is only important for very small grains
(VSG), polycyclic aromatic hydrocarbons (PAH), or other similar particles. The
overall result for a particular grain is excess emission at shorter wavelengths
(due to temperature spikes) and lower emission at longer wavelengths (due to
cold intervals). When a grain ensemble with an MRN-like size distribution
\citep{mrn} is considered, only excess mid-IR emission is seen
, as emission at longer wavelengths is dominated by large
grains that are not susceptible to stochastic heating.

The importance of stochastic heating is well recognized by the community dealing
with the galaxy-wide star formation. It now becomes a standard component of
galactic models to explain the emergent spectral energy distribution (SED)
\citep[e.g., ][]{Draine2007,Compiegne2011,Popescu2011,baes}. There are also models
of protostellar objects and circumstellar discs, which include the stochastic
heating \citep[e.g.,][]{HenningManske,cloudy,wood2008}. However, in studies of
individual protostellar objects this process is often ignored, and 24\,$\mu$m
emission is rather assumed to be an indication of the presence of an internal
heating source \citep[e.g.,][]{bs,rathborne2010}. This is probably justified in
cases when a compact source is seen on a 24\,$\mu$m image, but the
interpretation of diffuse 24\,$\mu$m emission can be less straightforward.

A common approach to the SED modelling of star-forming regions is an application
of the modified black-body law \citep{hildebrand1983}. This law is extensively
used in the analysis of far-infrared, submillimetre, and millimetre
observations. However, its application is limited only to the {\em emission\/}
of dust distributed along the line of sight. Accordingly, it only allows to find
the dust column density and density-weighted temperature, which is not enough
for the chemical modelling needed to interpret line observations. Also,
inferences, based on this law, can be inaccurate and ambiguous. For example,
observed variations in the dust emissivity index $\beta$ can be caused either by
real changes in dust properties, by observational noise, or by the temperature
gradient along the line of sight \citep{shetty1,shetty2}. 

Infrared Dark Clouds (IRDC), that are believed to be massive counterparts of
low-mass prestellar cores, are simultaneously seen in emission (millimetre and
submillimetre) and {\em absorption} (near-IR and mid-IR).  Thus, their proper RT
modelling hopefully makes the derived density and temperature structure less
ambiguous and more reliable.  Density and temperature distributions in a core
affect both the shape and the depth of the intensity minimum at shorter
wavelengths and the intensity maximum at longer wavelengths. Thus, to reproduce
the IRDC shadows and their millimetre emission counterparts, one needs a
spatially resolved, at least, 1D model of the RT in the cloud.

It needs to be taken into account that extended IR emission from an infrared
dark core consists of three components, namely, attenuated background emission,
proper core emission, and foreground emission. Proper core emission is caused by
dust grains heated both by an internal source and by the external radiation and
can be estimated using the core RT model. A major problem of infrared studies is
the problem of foreground subtraction. There are two possible approaches to this
problem. The first is to assume that the observed intensity at the darkest spot
of the considered region is entirely caused by foreground emission of any
nature, like zodiacal light, foreground interstellar matter, or instrument noise
\citep{stutz}. The second is to take the zodiacal light contribution from the
interplanetary dust cloud model \citep{dirbe1998}, then to subtract it from the
signal, and to assume that all the remaining intensity comes from the source
itself.

The first approach is safer as in this case one definitely knows that derived
results represent some limiting values. On the other hand, its usage implies an
assumption that the core is optically thick in the considered range and does not
produce any proper emission. These are two additional and maybe unwanted
constraints to the model. The second approach seems to be more honest, but it
brings up a question of a foreground subtraction.

The foreground problem can be related to the excess 24\,$\mu$m emission,
supposedly found in two IRDC cores by \citet{Pavlyuchenkov2011}. They attempted
to model SEDs of infrared dark cloud cores IRDC~320.27+029(P2) and
IRDC~321.73+005(P2) \citep{Vasyunina2009} in the range from 8\,$\mu$m to
1.2\,mm. \citet{Pavlyuchenkov2011} found that a model of a spherically symmetric
core, that only accounts for equilibrium dust thermal emission, allows to
reproduce absorption of background radiation at 8\,$\mu$m, emission (or the lack
of emission) at 70\,$\mu$m and 1.2\,mm, but fails at 24\,$\mu$m. Absorption
minima at this wavelength, predicted by the model, turned out to be much deeper
than is actually observed. \cite{Pavlyuchenkov2011} suggested that an excess
emission at 24\,$\mu$m can be generated by stochastically heated VSGs, that are
well-known potential sources of emission in the mid-IR range
\citep{Kruegel2003}.

In this paper we use an extended version of the RT model from
\citet{Pavlyuchenkov2011} to simulate emission of a dense interstellar clump
heated from outside by the interstellar UV radiation and possibly from inside by
a protostellar object. Emission from stochastically heated VSGs and PAHs is
taken into account. Our initial goal was to reproduce shallow 24\,$\mu$m shadows
in IRDC cores studied by \citet{Pavlyuchenkov2011}. While pursuing this goal, we
found that a stochastic dust heating model of a kind, that is used in galactic
SED modelling, being applied to individual clumps, predicts their distinct
morphological features. We discuss possible implications of this finding and
also consider a role of other non-radiative dust heating processes.

The structure of the paper is the following. In Section~2 we discuss the problem
of foreground emission and show how its level of uncertainty affects the derived
core parameters. The protostellar core model and the RT method with stochastic
heating algorithm are described in Section~3. Also, in this section results of
protostellar core simulations with stochastically heated dust grains are
presented. In Section~4 we consider possible ways to solve the excess mid-IR
emission problem. Our conclusions are summarized in the last section.

\section{Continuous dust heating and the foreground emission}

As we mentioned in the introduction, there are two ways to tackle the foreground
problem. One of them is to assume that the optical depth at the darkest spot of
the studied region is so large that all the emission from this location  is
caused by foreground sources. The other approach is to estimate the foreground
emission as a sum of the zodiacal light \citep{dirbe1998} and the interstellar
matter contribution \citep{bt2009}. The two approaches may lead to quite
different inferences.

{ Let us take the IRDC~321.73+005 (P2) core (IRDC~321 for short) from
\cite{Pavlyuchenkov2011} as an example.} The background intensity at 24\,$\mu$m
in its vicinity is about 35\,MJy/ster, while the intensity at the peak of the
millimetre emission is less than 30\,MJy/ster. The zodiacal light contribution
{ at 24\,$\mu$m}, as given by the SPOT software, is about 18\,MJy/ster. As
this object is relatively nearby \citep[$\sim2$ kpc,][]{Vasyunina2009}, the
smoothed Galactic foreground contribution, estimated as in \cite{bt2009}, is
less than 10\%, so we neglect it here. Subtracting the zodiacal light only, we
get $\sim17$\,MJy/ster for the background emission (plus any unaccounted
foreground emission) and 10\,MJy/ster for the emission at the millimetre peak.
These are intensities that have been used by \cite{Pavlyuchenkov2011}.
Alternatively, if we would assume that all the extra emission at the bottom of
the shadow is the foreground emission, we would end up with zero emission for
the core and $\sim5-7$\,MJy/ster for the surrounding region.

To illustrate the significance of the choice, we repeat here calculations from
\cite{Pavlyuchenkov2011} using these two approaches to the foreground estimation.
Prior to giving results we recall the basic equations of the adopted RT model
with continuous dust heating. The core is assumed to be spherically symmetric,
with the density distribution given by the same expression as is used for
low-mass prestellar cores \citep{Tafalla2002}:
\begin{equation}
n({\rm H}_2) = \frac{n_0}{1+(r/r_0)^p},
\end{equation}
where $n_0$ is the central density, $r_0$ is the radius of an inner plateau, $p$
describes the density fall-off in the envelope. To take into account the heating
by an embedded protostar (or a group of protostars) we assume that there is a
black-body source in the core centre with the temperature $T_*$ and a fixed
radius of 5\,$R_\odot$. The inner core radius is set to 50\,AU (which is the
upper limit for the dust sublimation radius), and the outer core radius is
assumed to be equal to 1\,pc. The core is illuminated by the diffuse ambient
interstellar field that is characterized by a certain colour temperature 
$T_{\rm bg}$ and dilution $D_{\rm bg}$.

{ We use the Accelerated Lambda Iteration (ALI) method for the radiative
transfer modelling, which  is similar to that described
in~\cite{Pavlyuchenkov2004} and in \cite{hoger}, but with modifications for
thermal radiation \citep[see also][for general principles of the ALI]{hubeny}.}
The mean radiation intensity
\begin{equation}
J_\nu=(4\pi)^{-1}\int\limits_{4\pi}I_{\nu}(\vec{n})d\Omega
\end{equation}
is determined by integrating the radiative transfer equation
\begin{equation}
\left(\vec{n}\nabla\right)I_{\nu} = \kappa_{\nu}\left(S_{\nu} -I_{\nu}\right)
\label{radtran}
\end{equation}
along representative directions. Here $I_{\nu}$ is the specific radiation intensity,
$S_\nu$ is the source function, $\kappa_{\nu}=\alpha_{\nu}+\sigma_{\nu}$ is
the extinction coefficient, $\alpha_{\nu}$ is the absorption coefficient, and
$\sigma_{\nu}$ is the scattering coefficient.
The temperature of the medium $T$ is found from the equation of radiative equilibrium
\begin{equation}
\int\limits_{0}^{\infty}\alpha_{\nu}J_{\nu}d\nu =
\int\limits_{0}^{\infty}\alpha_{\nu}B_{\nu}(T)d\nu,
\end{equation}
where $B_{\nu}$ is the Planck function. { The adopted convergence criterion
is that the relative difference in temperature between subsequent iterations is
smaller than $10^{-3}$ at all radii. Although this formalism is
multidimensional, in this study the RT equation is solved in 1D, under the
assumption of spherical symmetry.}

In the radiative transfer equation, scattering is taken into account in the
approximation of isotropic coherent scattering, so that the source function
$S_{\nu}$ takes the form
\begin{equation}
S_{\nu}=\dfrac{\alpha_{\nu}B_{\nu}+\sigma_{\nu}J_{\nu}}{\kappa_{\nu}}.
\end{equation}
The absorption and scattering coefficients as functions of frequency for
amorphous silicate grains of a given radius are computed using the Mie theory.
To calculate the total absorption and scattering coefficients for a dust
ensemble we assume that the size distribution of dust grains is described by a
power law, $f(a)\propto a^{-3.5}$ \citep{mrn}, with minimum and maximum grain
radii of 0.001 and 10 $\mu$m. { For each parameter combination of the core
model, we simulate the radiative transfer, compute the intensity distributions at
each considered wavelength, and quantitatively compare the computed and observed
distributions using the standard $\chi^2$ criterion. The theoretical
distributions are convolved with the relevant telescope beams for each
wavelength.  The search for best-fit parameter values is performed with the
PIKAIA genetic algorithm \citep{pikaia}. More details about the fitting
procedure can be found in \cite{Pavlyuchenkov2011}.}

\begin{figure*}
\includegraphics[width=0.4\textwidth,clip=]{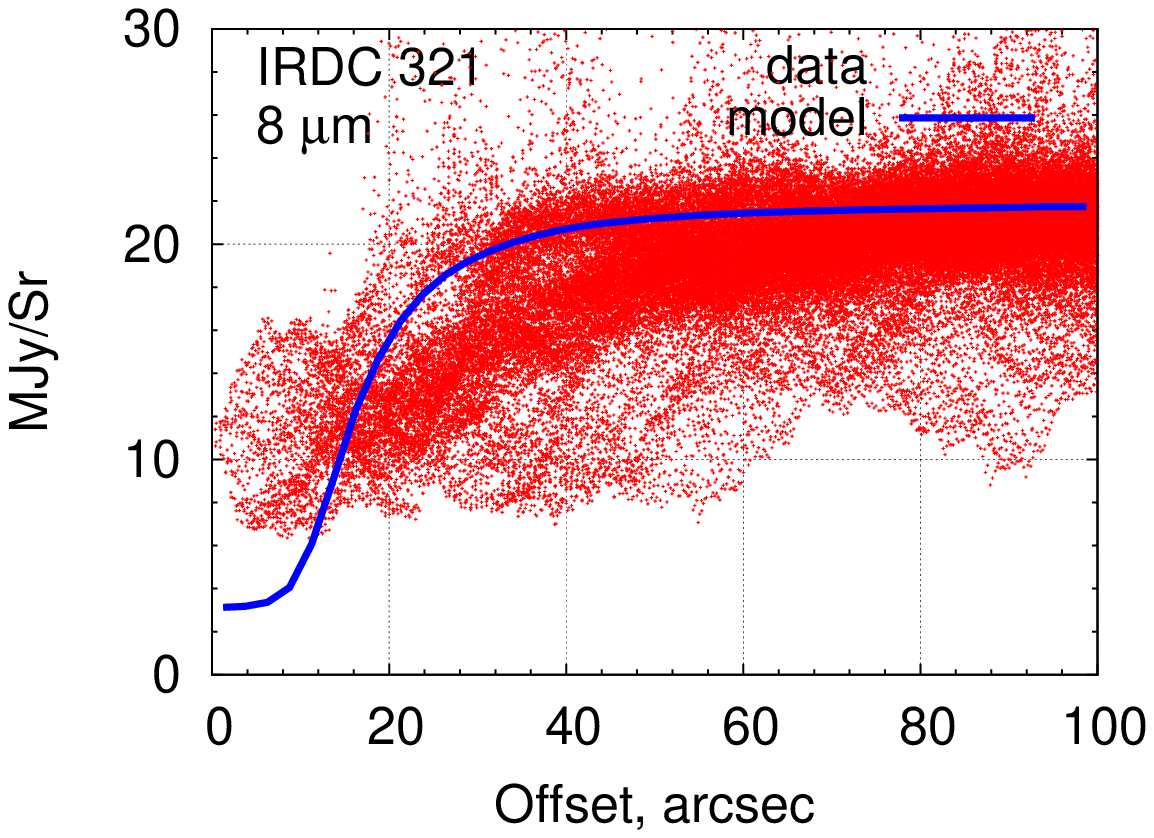}
\includegraphics[width=0.4\textwidth,clip=]{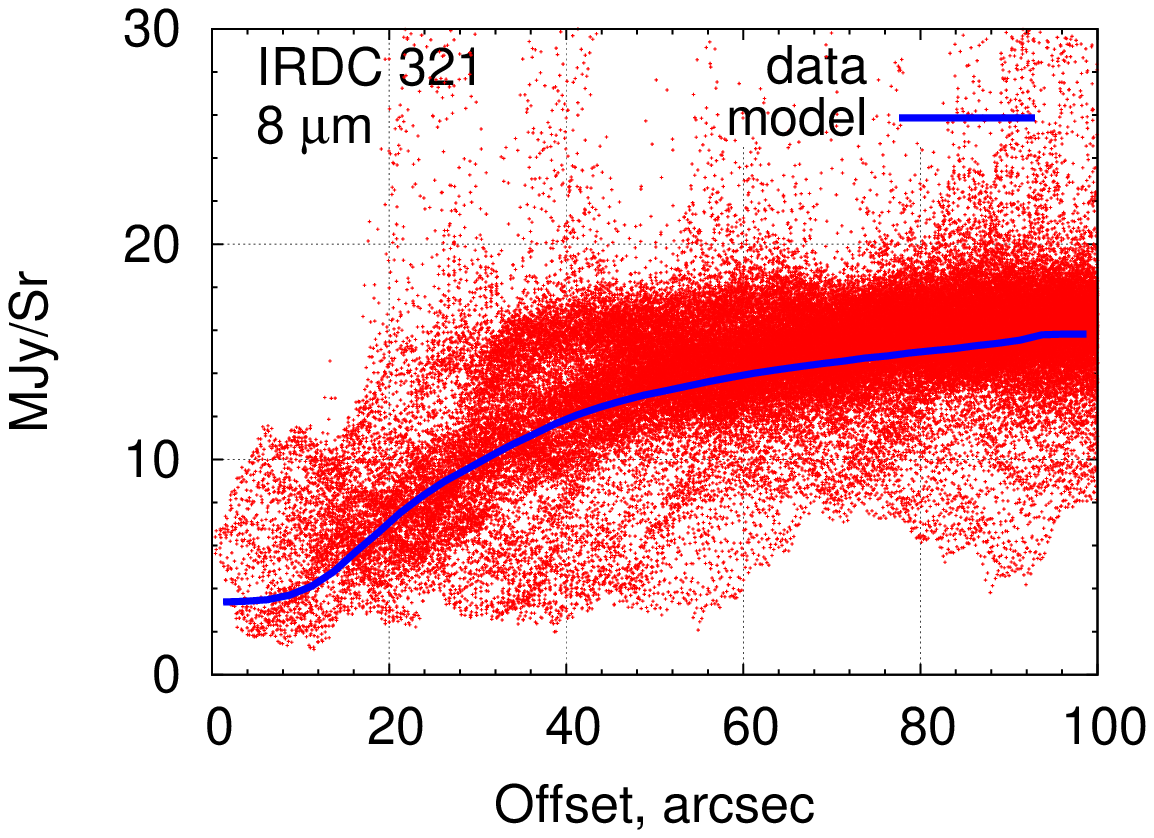}

\includegraphics[width=0.4\textwidth,clip=]{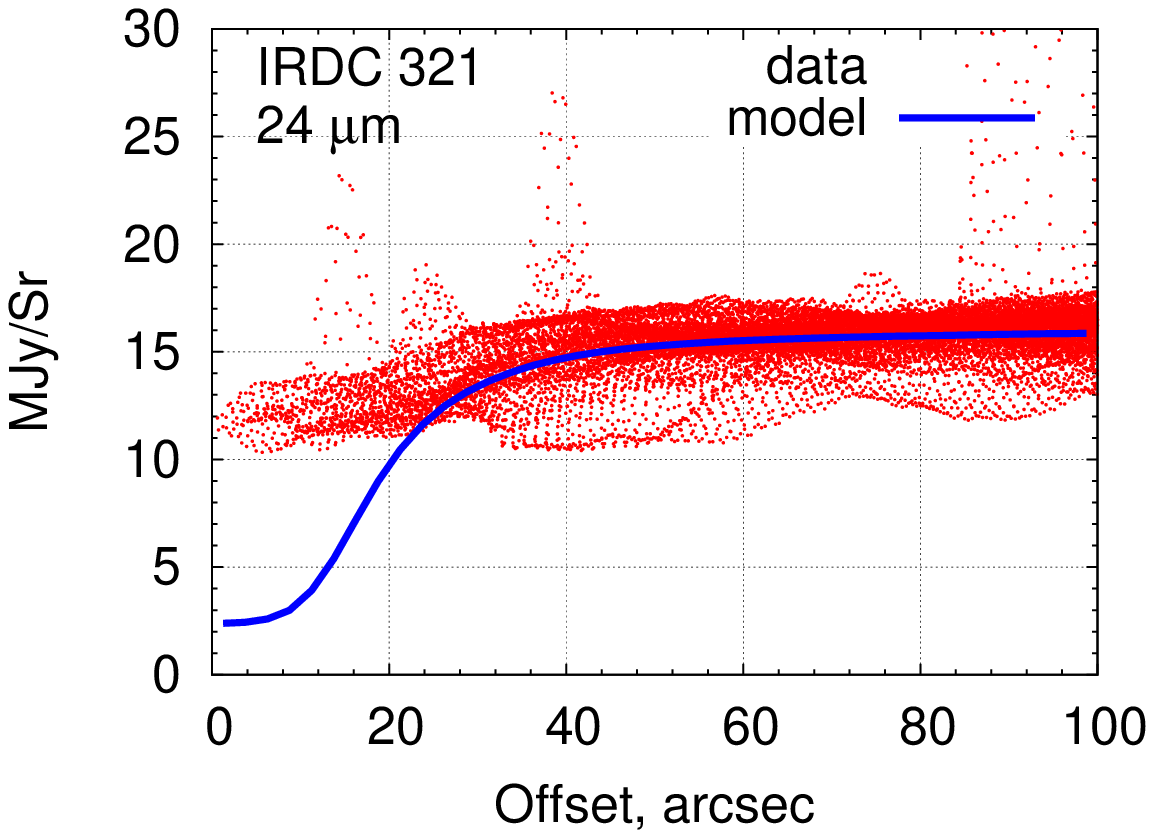}
\includegraphics[width=0.4\textwidth,clip=]{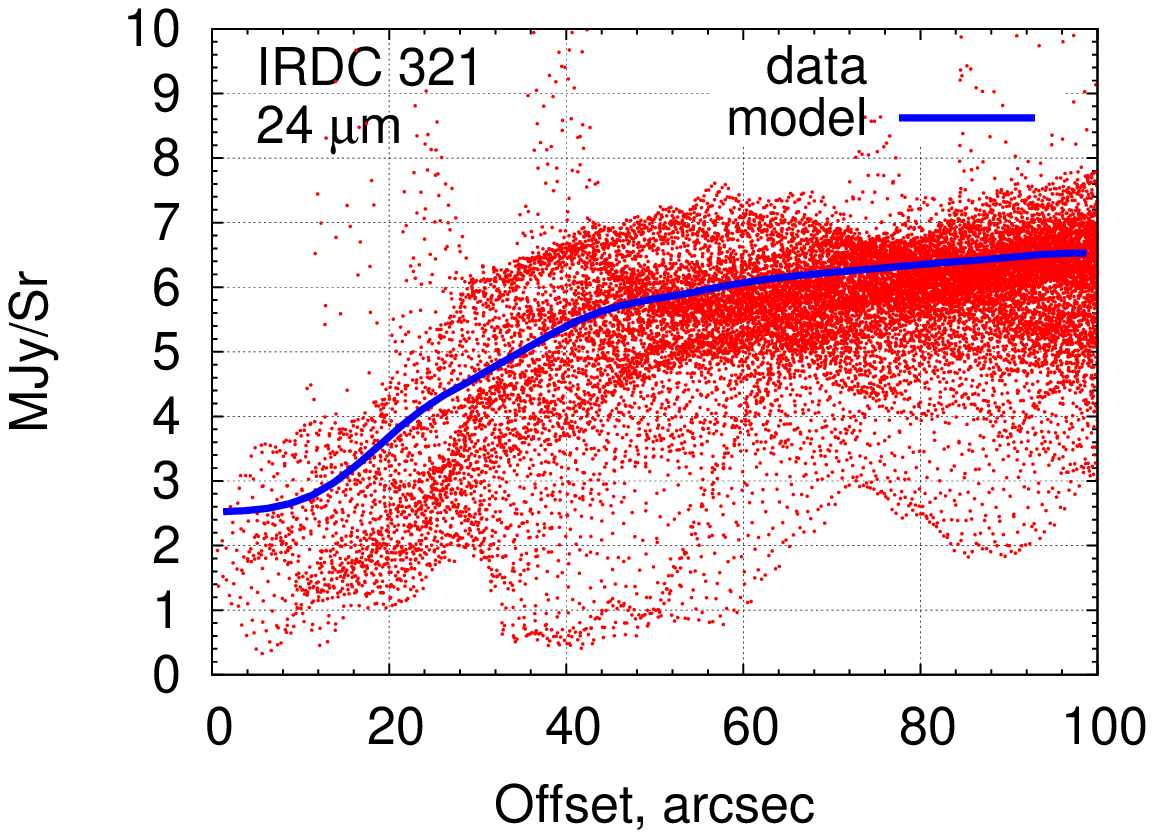}

\includegraphics[width=0.4\textwidth,clip=]{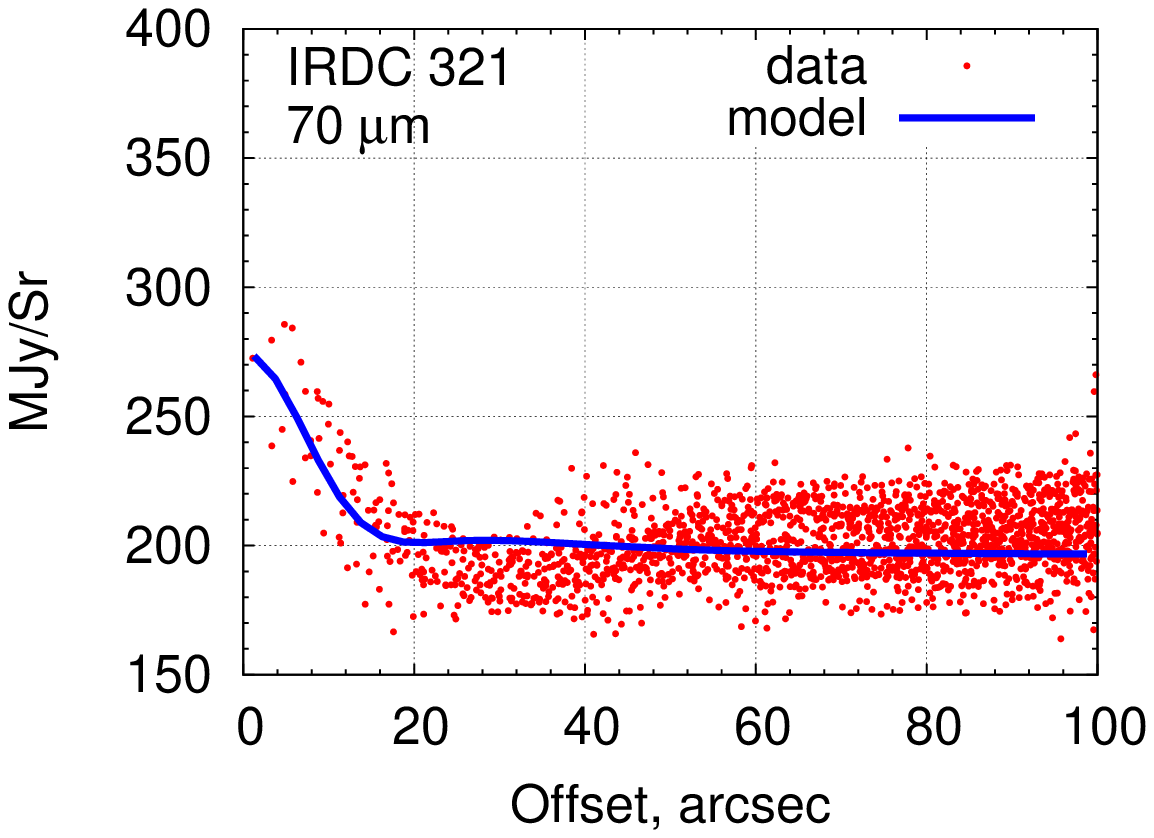}
\includegraphics[width=0.4\textwidth,clip=]{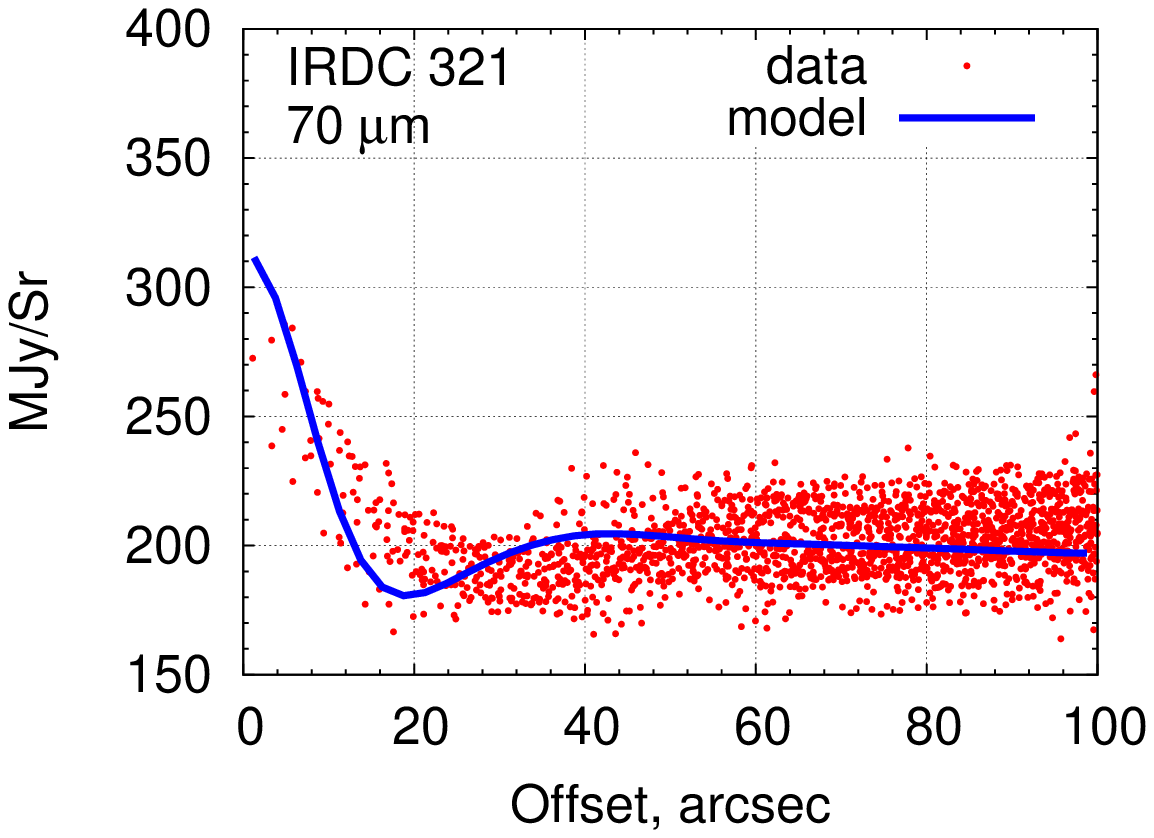}

\includegraphics[width=0.4\textwidth,clip=]{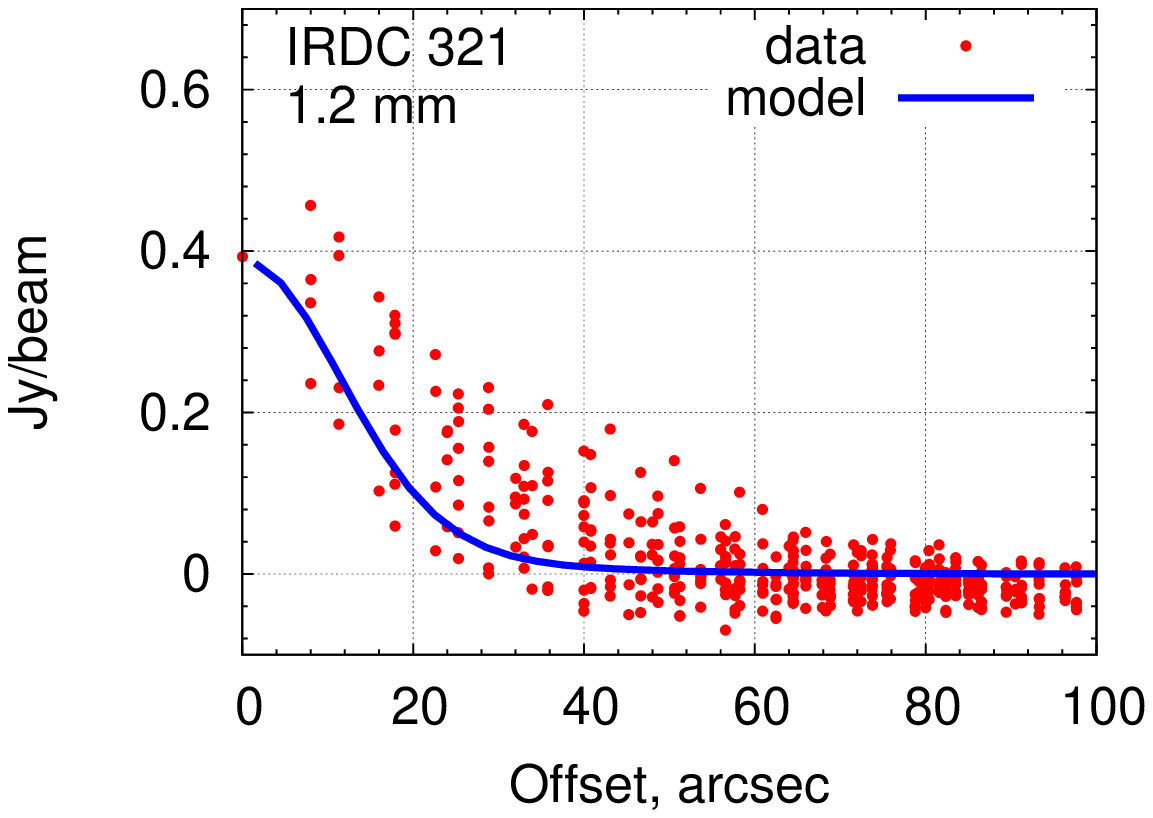}
\includegraphics[width=0.4\textwidth,clip=]{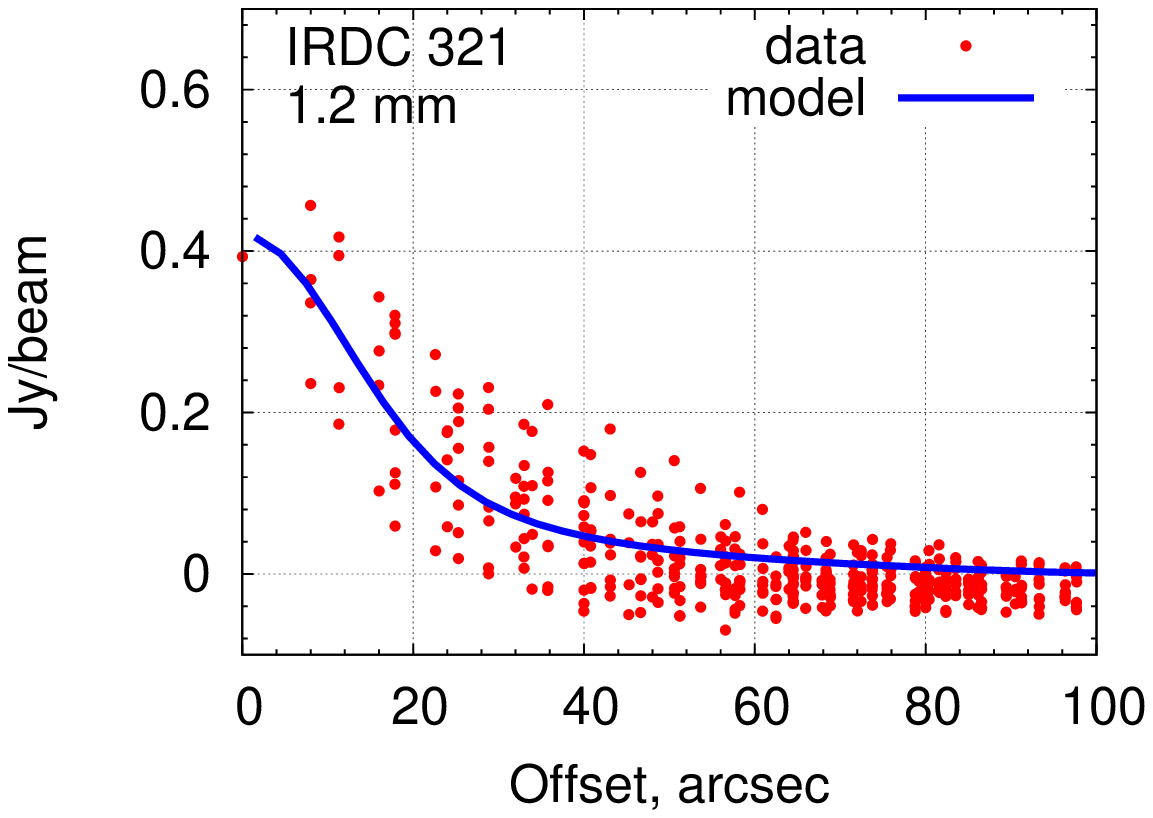}

\caption{Theoretical and observed integrated intensities in the IRDC~321 core
computed with different assumptions on the foreground. In the left column all
the foreground emission is caused by the zodiacal light estimated using the
DIRBE model (DZF). In the right column it is assumed that the foreground
emission intensity is equal to the intensity at the darkest region of the core
(CFS). { Dots represent individual observational pixels}.}
\label{backminus}
\end{figure*}

Results of the fit are shown in Figure~\ref{backminus}. In the left column we
present data for the case when only DIRBE-based zodiacal light estimate is
subtracted from the signal (hereinafter DZF case). { Practically, we subtract
zodiacal light from 8\,$\mu$m and 24\,$\mu$m data. Its contribution at
70\,$\mu$m and 1.2\,mm is negligible.}  The right column in
Figure~\ref{backminus} corresponds to the case when the foreground level is
assumed to be equal to the lowest intensity within the IRDC~321 core
(hereinafter complete foreground subtraction, CFS case).

\begin{figure}
\includegraphics[width=0.4\textwidth,clip=]{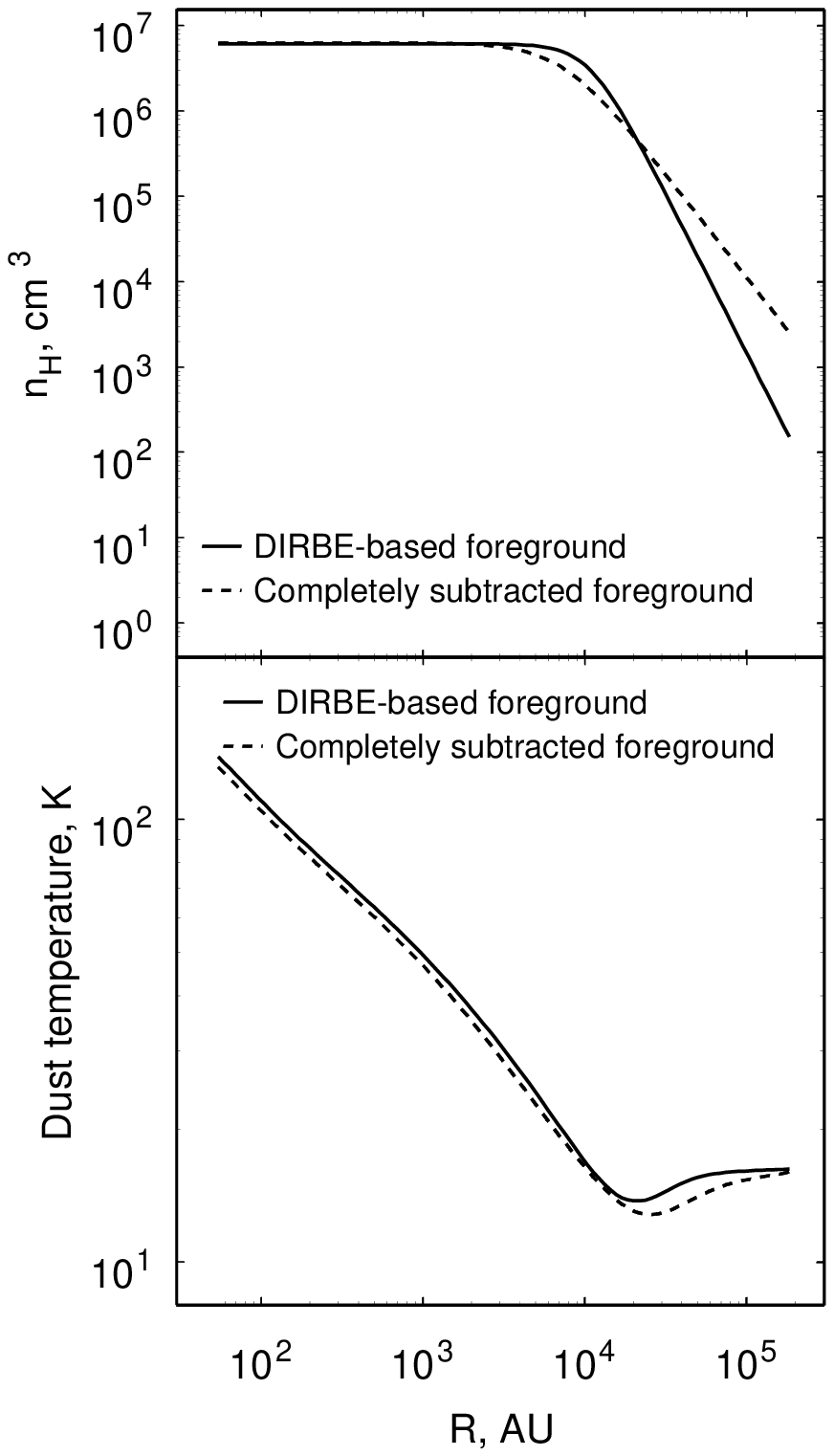}
\caption{Density and temperature radial profiles for the models presented in
Figure~\ref{backminus}.}
\label{struct}
\end{figure}

Obviously, in the CFS case we do not have problems either with 24\,$\mu$m emission
or with other wavelengths. At the same time, some core parameters, that are
derived under this assumption, are quite different from the results obtained in
the DZF case. While temperature profiles and central densities $n_0$ are almost
identical, slopes of the power-law envelope $p$ are markedly different
(Figure~\ref{struct}). In the CFS model $p=2.5$, which is typical for prestellar
objects. In the DZF model the density fall-off is much steeper, with $p=3.7$.
Apparently, in the DZF case the algorithm tried to reconcile the need to have
more material to account for emission at longer wavelengths and the need to have
less material to account for low absorption at shorter wavelengths. As a result,
the core mass is 310\,$M_\odot$ in the DZF model and 750\,$M_\odot$ in the CFS
model. One may argue that the factor of two difference in the core mass is not that
significant. However, different slopes of the density radial profile in the
envelope may lead to a greater effect when applied to chemical modelling and
molecular line RT modelling.

We should note that in general the CFS model does
look more attractive. It produces more physical density profile, is consistent
with the 24\,$\mu$m intensity radial distribution. Also, analysis of the
algorithm convergence shows that the solution in the CFS case is well-defined,
while in the DZF case an alternative solution exists with higher central density
($\sim10^9$\,cm$^{-3}$) and smaller plateau \citep{Pavlyuchenkov2011} which is
nearly as good (by formal $\chi^2$ criterion) as the low density solution
presented here. 

So, after all, to reproduce all the bands, it seems that it is only necessary to
assume that all the extra emission comes from the foreground, i.e. to rely on
the CFS approach. But how justified is this assumption? Let us consider some
general arguments. The distance to the IRDC 321 core that we use as an example
is about 2\,kpc. To account for the 24\,$\mu$m emission, as described above, we
would need an interstellar foreground fraction as high as 70\%. However, an
approximate method described in \cite{bt2009} gives the expected foreground
contribution of only about 10\% for this object. Also, according to the Galaxy
structure map, based on the GLIMPSE survey \citep{glimpse}, there should not be
much intervening material in the direction of the IRDC 321 core.

\begin{figure}
\psfig{figure=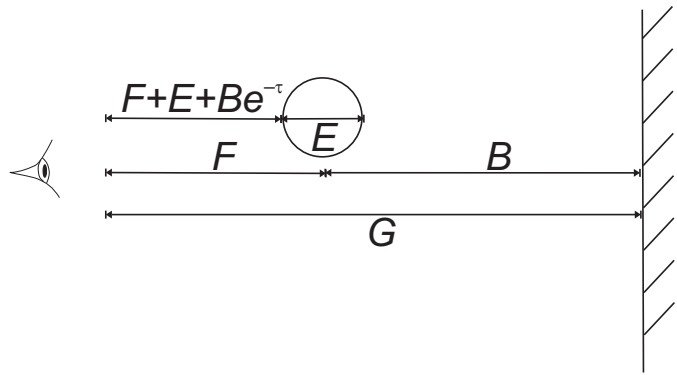,width=0.5\textwidth}
\caption{Scheme explaining various symbols entering equations (\ref{eta1}) and (\ref{eta2}).}
\label{galc}
\end{figure}

\begin{figure}
\psfig{figure=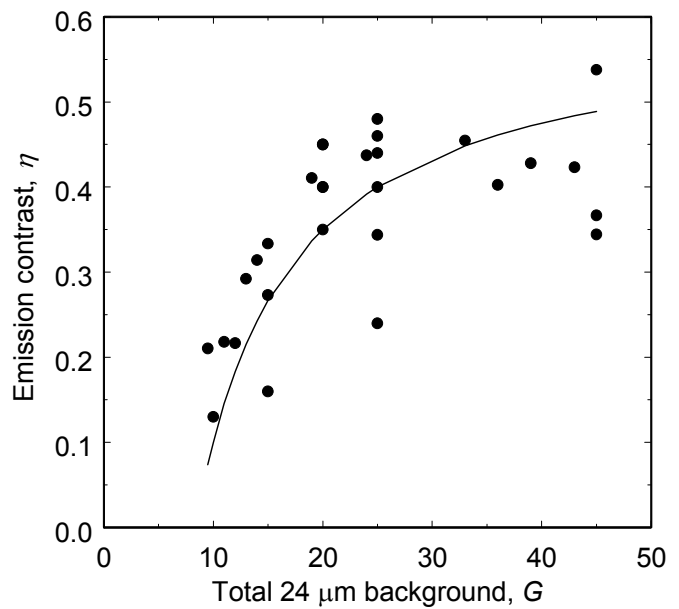,width=0.5\textwidth}
\caption{Emission contrast $\eta$ at 24\,$\mu$m for cores from samples by Vasyunina et al. (2009)
and Butler \& Tan (2009) as a function of the total IR background intensity.
Zodiacal light contribution is subtracted.}
\label{contrast}
\end{figure}

Let us assume that all IRDC cores are identical, that is, they all have the same
optical depth $\tau$ and emissivity $E$ (at a certain wavelength). Also, let us
assume that the observed emission $G$ outside of the core projection is the sum
of the `true' background emission $B$ (that comes from the material behind the
core) and the foreground emission $F$, that constitutes the fraction $f$ of the
total emission, $F=fG$ (Figure~\ref{galc}). Then, the observed emission contrast
\begin{equation}
\eta={G-(Be^{-\tau}+E+F)\over G}.
\label{eta1}
\end{equation}
After some algebra we obtain
\begin{equation}
\eta=(1-f)(1-e^{-\tau})-{E\over G}.
\label{eta2}
\end{equation}
In Figure~\ref{contrast} we show the observationally inferred $\eta$ for cores
from \cite{Vasyunina2009} and \cite{bt2009} samples at 24\,$\mu$m { (after
zodiacal light subtraction)}. As we do not
intend to make any precise conclusions, values are simply read off by eye from
the MIPSGAL cutouts. The plot apparently shows that $\eta$ grows for $G<30$\,MJy
ster$^{-1}$ and then stays constant. If the proper core emission were absent, that
is, $E=0$, we would have $\eta$ independent of $G$ (of course, under the
assumption that $f$ is independent of $G$). On the other hand, if $E$ is not
zero, $\eta$ would be an increasing function of $G$ as long as $G$ is not too
big to make the second term small. This is really observed for $G<30$\,MJy
ster$^{-1}$. The increase of $\eta$ by 0.4 from $G=10$\,MJy ster$^{-1}$ to
$G=30$\,MJy ster$^{-1}$ corresponds to $E\approx5$\,MJy ster$^{-1}$. At larger
values of $G$ the second term in eq.~(\ref{eta2}) is small, and $\eta$ stays
nearly constant with some scatter caused by scatter in $f$ (that is, in
distances and Galactic structure features).


Solid line in Figure~\ref{contrast} is drawn using eq.~(\ref{eta2}) for $f=0.4$
and $E=5$\,MJy ster$^{-1}$ ($\tau_{24}=\infty$). Thus, observed $\eta$ values
indicate that there exists a typical foreground fraction $f\sim0.4$, that is
close to the value of 0.54 inferred by \cite{PerettoFuller2009}. { This is a
statistical value, of course, that is not directly applicable to specific
objects. For example, in Figure~\ref{contrast} a point corresponding to IRDC~321
lies above the solid line, which is consistent with $f$ being somewhat smaller
than 0.4.} An analogous plot for an 8\,$\mu$m contrast does not show any clear
correlation with $G$, with an average $\eta_8$ value of 0.6 and significant
scatter. This also implies $f\sim0.4$ (assuming $E_8=0$ and $\tau_8=\infty$).

So, the observed IRDC 24\,$\mu$m contrasts are broadly consistent with the
omnipresence of the intrinsic mid-IR emission in IRDC cores. In other words, the
DZF approach can actually be valid, indicating, though, that we miss some
important mechanism(s) responsible for the intrinsic mid-IR emission of the
cores.

It is important to remember that different assumptions on the foreground lead to
different inferences on the core structure, mass, chemical composition, and, by
implication, on its evolutionary status. If we intend to use the information,
deduced from the continuum emission modelling, to simulate molecular line
profiles, we do need a more substantiated approach to the foreground problem. In
the following section we check whether the discrepancy in the 24\,$\mu$m
emission can be alleviated by taking into account the stochastic heating of
small dust grains that is known to affect a SED in the mid-IR range.

\section{Stochastic Dust Heating}
{ To include effects of stochastic dust heating, the RT model needs to be
upgraded. First, it is no longer possible to assume that all dust grains have
the same temperature irrespective of their size. Second, a small grain is not in
thermal equilibrium with the radiation field, and its temperature varies with
time. So, in general, dust temperature of a single particle is a function of
grain size and time.

To keep the problem tractable, we drop the continuous size distribution and
assume that dust consists of finite number of components, $N$,} so absorption,
scattering and emission coefficients are given by
\begin{equation}
\alpha_{\nu}=\sum_{i=1}^{N} \alpha_{\nu}^{i};\hspace{0.5cm}
\sigma_{\nu}=\sum_{i=1}^{N} \sigma_{\nu}^{i};\hspace{0.5cm}
j_{\nu}=\sum_{i=1}^{N} j_{\nu}^{i}.
\end{equation}
Individual absorption and scattering coefficients are given by
\begin{align}
&\alpha_{\nu}^{i} = n_{i} \pi a_i^2 Q_{\nu,i}^{\rm abs} \\
&\sigma_{\nu}^{i} = n_{i} \pi a_i^2 Q_{\nu,i}^{\rm sca},
\end{align}
where $n_i$, $a_i$, $Q_{\nu,i}^{\rm abs}$, and $Q_{\nu,i}^{\rm sca}$ are the
number density, radius, absorption and scattering efficiency factors for dust
grains of $i$th type.

Since an isolated dust particle is subject to temperature fluctuations, we
consider a large ensemble of identical particles and describe their thermal
state by temperature probability density distribution $P(T)$. The $P(T)dT$ is
the fraction of particles with temperatures lying in the interval $(T,T+dT)$.
The emission coefficient for each dust component is given by
\begin{equation}
j_{\nu}^{i}=\alpha_{\nu}^{i}\int\limits_{0}^{\infty} P^{i}(T) B_{\nu}(T)\,dT,
\end{equation}
where $P^{i}(T)$ is the probability density distribution for $i$th component,
$B_{\nu}(T)$ is the Planck function. Distributions $P^{i}(T)$ are supposed to be
known from the previous iteration.

{ Similarly to the RT model with continuous heating, outlined in the previous
section, the method with stochastic dust heating is also based on the ALI
technique and consists of two steps. At the first step, the mean radiation
intensity is computed at each cell by integrating the RT equation
\eqref{radtran} in 1D along representative directions.}
At the second step, the mean intensity $J_{\nu}$ is used to update $P^{i}(T)$
for all dust components. Evaluation of $P^{i}(T)$ is based on the Monte Carlo
method. The temperature evolution of an isolated dust grain is simulated and
then converted into the probability density distribution. A single grain
temperature evolution is computed in three steps:
\begin{itemize}
\item generation of time and frequency sequences of absorbed photons;
\item calculation of grain temperature spikes due to absorption of photons; 
\item calculation of continuous grain cooling due to thermal emission between two consecutive
absorption events.
\end{itemize}
More details about the stochastic heating algorithm can be found in Appendix.

In this study we assume that dust consists of four components: large silicate
grains, large graphite grains, small graphite grains, and PAH particles. Their
parameters are listed in Table~\ref{dustparam}. Absorption and scattering
efficiency factors $Q_{\nu}^{\rm abs}$ and $Q_{\nu}^{\rm sca}$ for silicate and
graphite particles are taken from \cite{Laor1993}. Optical properties of PAHs
are calculated following~\cite{Draine2007}, where 30 PAH features (described by
the Drude profile) are taken into account. Heat capacities $C_{V}(T)$ are taken
from~\cite{Draine2001}.
\begin{table*}
\caption{Dust parameters.}
\label{dustparam}
\begin{tabular}{llllll}
\hline
Component & Radius, & Heat capacity pa-      & Density,  & Atomic & Mass \\
          & cm      & rameter $T_{\rm d}^{*}$, K  & g cm$^{-3}$ & mass & fraction \\
\hline
Large silicate grains & $3\cdot10^{-5}$ & 175 & 3.5 & 28 & 0.70 \\
Large graphite grains & $2\cdot10^{-5}$ & 450 & 1.81& 12 & 0.15 \\
Small graphite grains & $3\cdot10^{-7}$ & 450 & 1.81& 12 & 0.05 \\
PAHs                  & $7\cdot10^{-8}$ & 450 & 2.24& 12 & 0.10 \\
\hline
$^{*}$ See Appendix for the definition of $T_{\rm d}$.
\end{tabular}
\end{table*}
{ Simulations of IRDCs are performed using 100 logarithmically spaced radial
cells. As we use the Monte Carlo method to compute $P(T)$, the resultant
distributions are quite noisy. So we adopt a much softer approach here, assuming
5\% difference in $P(T)$ between subsequent iterations as a stopping criterion.
Each model requires about 5 Lambda-iterations for convergence. The total
computation time is about one day per model on an average PC. }

The code was tested against a number of problems. In the case of large grains
(when stochastic heating is not important) results are consistent with
temperatures computed using the code with continuous dust heating. The
stochastic heating algorithm was tested separately. In particular, we
successfully reproduce SED and $P^{i}(T)$ distributions shown in Figures~2 and~6
from \cite{Popescu2011}.

\subsection{Stochastic heating by interstellar UV radiation}

For the computations below, density distribution and the inner source parameters
are those from the best-fit IRDC 321 model from \cite{Pavlyuchenkov2011} with
the DIRBE-based zodiacal background subtraction. The diffuse interstellar UV
field is represented by a Planck spectrum with $T_{\rm bg}=20000$\,K and
dilution of $10^{-16}$.

\begin{figure*}
\centerline{\psfig{figure=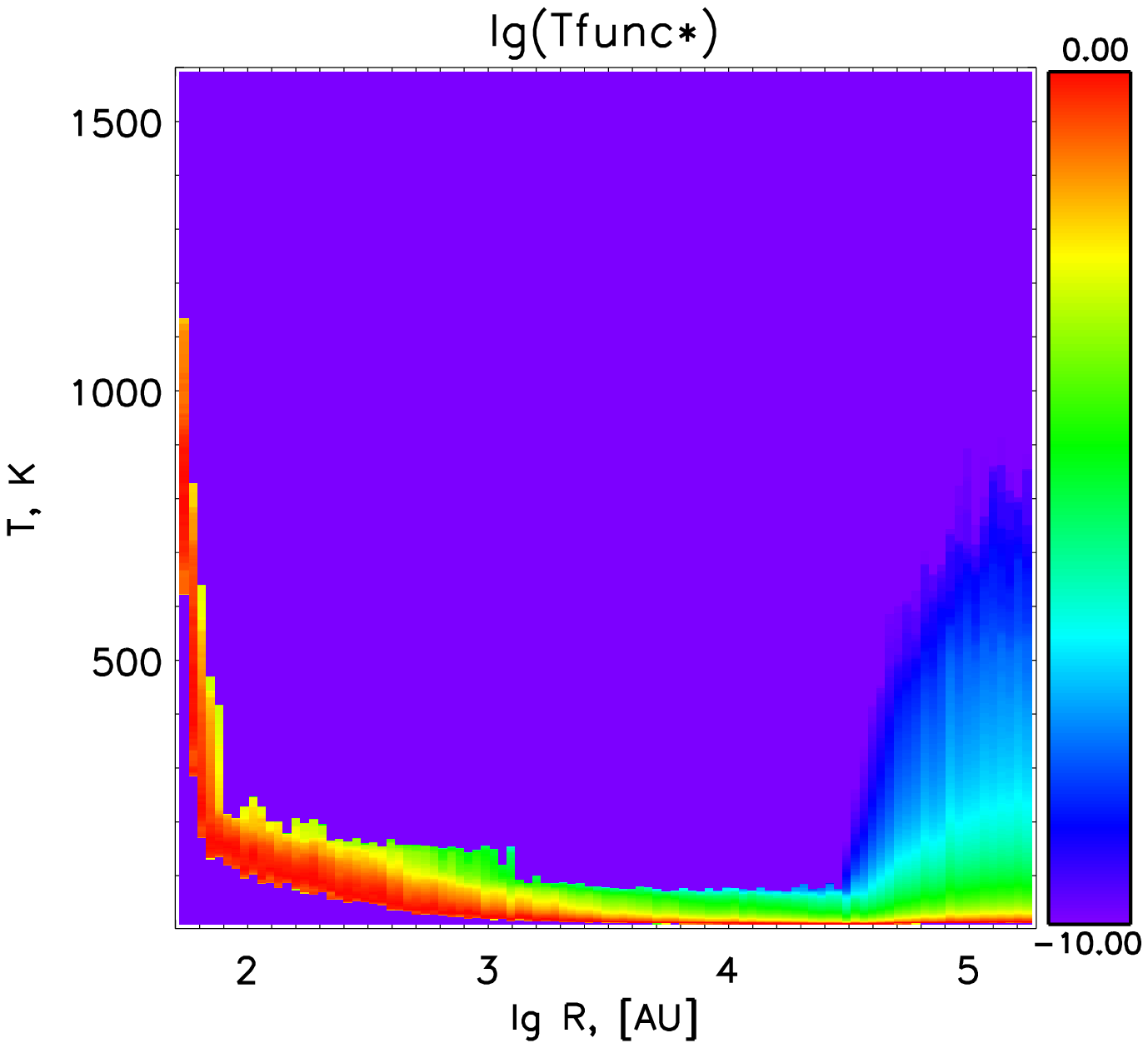,width=0.45\hdsize}
\psfig{figure=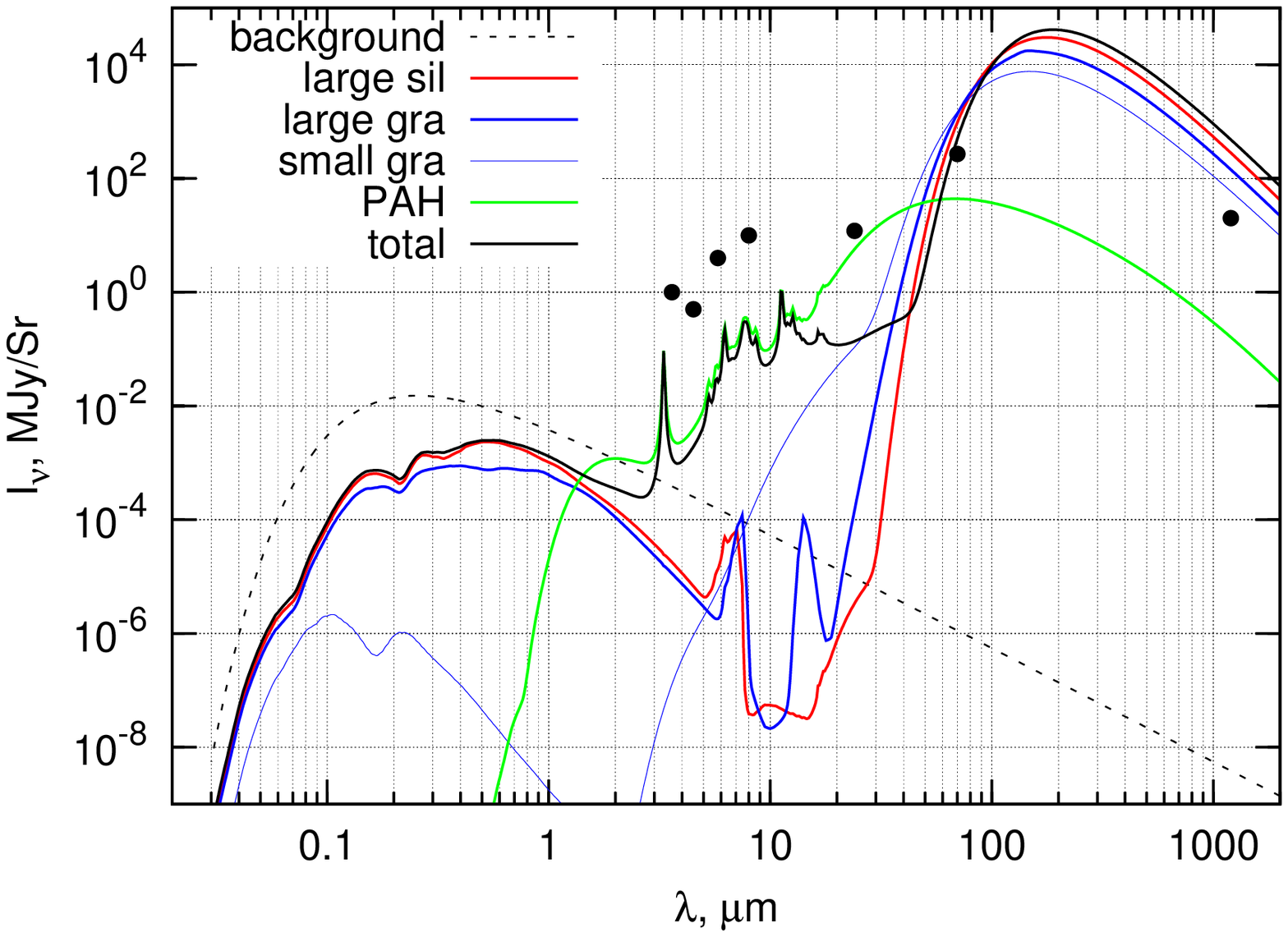,width=0.55\hdsize}}
\caption{Results of RT simulations for the model with the standard interstellar
field. Left: shown with colour is logarithm of the normalized temperature
probability distribution $P^{*}(T)$ for PAH particles ($P^{*}(T) =
P(T)/\max\{P(T)\}$). Right: spectral energy distribution toward the position
offset by 1$^{\prime\prime}$ from the core centre. Spectra are computed
separately for large silicates (red lines), large graphite grains (thick blue
lines), small graphite grains (thin blue lines), and PAH particles (red lines).
The emergent spectrum with all dust populations taken into account is shown with
solid black line. The background interstellar radiation is shown with black
dashed line. Filled circles indicate observed values for the IRDC 321 core.}
\label{Tfuncnorm}
\end{figure*}

In Figure~\ref{Tfuncnorm} (left) we present radial dependence of the normalized
probability density distribution $P(T)$ for PAH particles ($P(T)$ for VSGs looks
similar). The spread of temperatures in the core envelope is very broad. While
at any given time most of PAH particles are cold (20\,K), their small fraction
is heated up to 1000\,K. Maximum temperatures near the protostar also exceed
1000\,K. In the rest of the core $P(T)$ distribution is relatively narrow,
corresponding to the equilibrium temperature.

The SEDs for the model core toward the position, offset by 1$^{\prime\prime}$
from the core centre (to exclude the direct emission from the central source),
are shown in Figure~\ref{Tfuncnorm} (right). No convolution with a telescope
beam is applied { in this and subsequent SEDs. It is hard to isolate the
relative contribution from various dust components on a single spectrum. To show
their roles, after simulating the thermal structure of the object with all four
components simultaneously, we computed the emergent spectra separately for each
component. More precisely, at the ray-tracing stage we only took into account
emission, absorption, and scattering by a single dust component. These
contributions are shown in right panel of Figure~\ref{Tfuncnorm} with coloured
lines. With the solid black line we show the combined spectrum computed for all
the four dust populations.}

\begin{figure}
\centerline{\psfig{figure=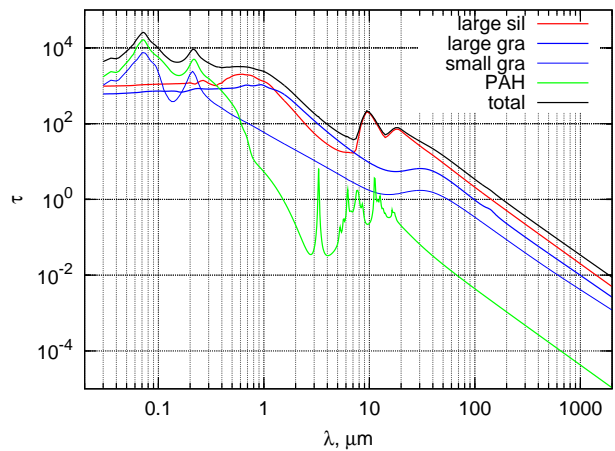,width=0.5\hdsize}}
\caption{
Optical depths { through the entire core} toward the position, offset by 1$^{\prime\prime}$
from the core centre, for various dust components.}
\label{fig_tau}
\end{figure}

Physically, the spectrum can be divided into three intervals, according to the
optical depth shown in Figure \ref{fig_tau}. At mm and sub-mm wavelengths
thermal emission from silicate and graphite grains dominates the spectrum. The
optical depth at these frequencies is low (Figure~\ref{fig_tau}), and
intensities depend significantly on details of the thermal structure in the
interior of the core, in particular, on the presence or absence of a protostar.
At the IR range (1--24\,$\mu$m) thermal emission from stochastically heated PAHs
is mostly responsible for the emergent radiation. The total optical depth at
these wavelengths is relatively high, and the spectrum is not sensitive to the
presence or absence of the internal heating source. At even higher frequencies
($\lambda<1\,\mu$m) the emergent spectrum is formed by large silicate grains via
scattering of background radiation. The optical depth here is very high, and the
spectrum is not affected by the interior of the core.

In Figure~\ref{Tfuncnorm} (right) intensities observed toward the centre of the
IRDC 321 core are indicated by filled circles. Obviously, we do not solve the
mid-IR intensity problem even by adding the stochastic heating effects. More
accurately, if there were only PAHs in the core, combined emission of
stochastically heated particles in the core envelope and in the vicinity of the
protostar would provide the needed intensity at 24\,$\mu$m. However, in the
model with all four dust components the emission of PAHs from the core centre is
completely absorbed by intervening large grains. At shorter wavelengths even
combined (core+envelope) PAH emission is not sufficient to explain observed
intensity.

\begin{figure*}
\centerline{\psfig{figure=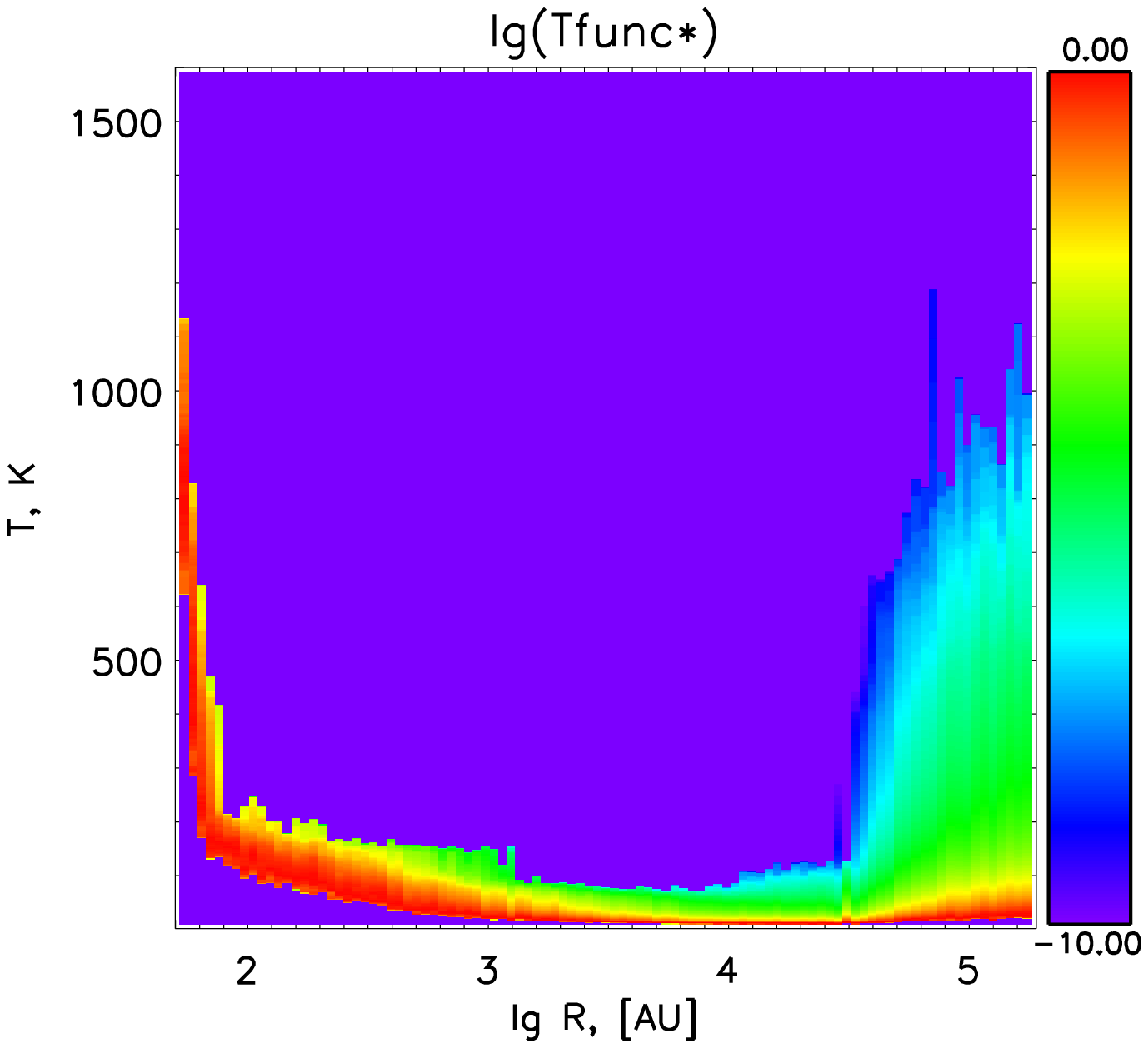,width=0.45\hdsize}
\psfig{figure=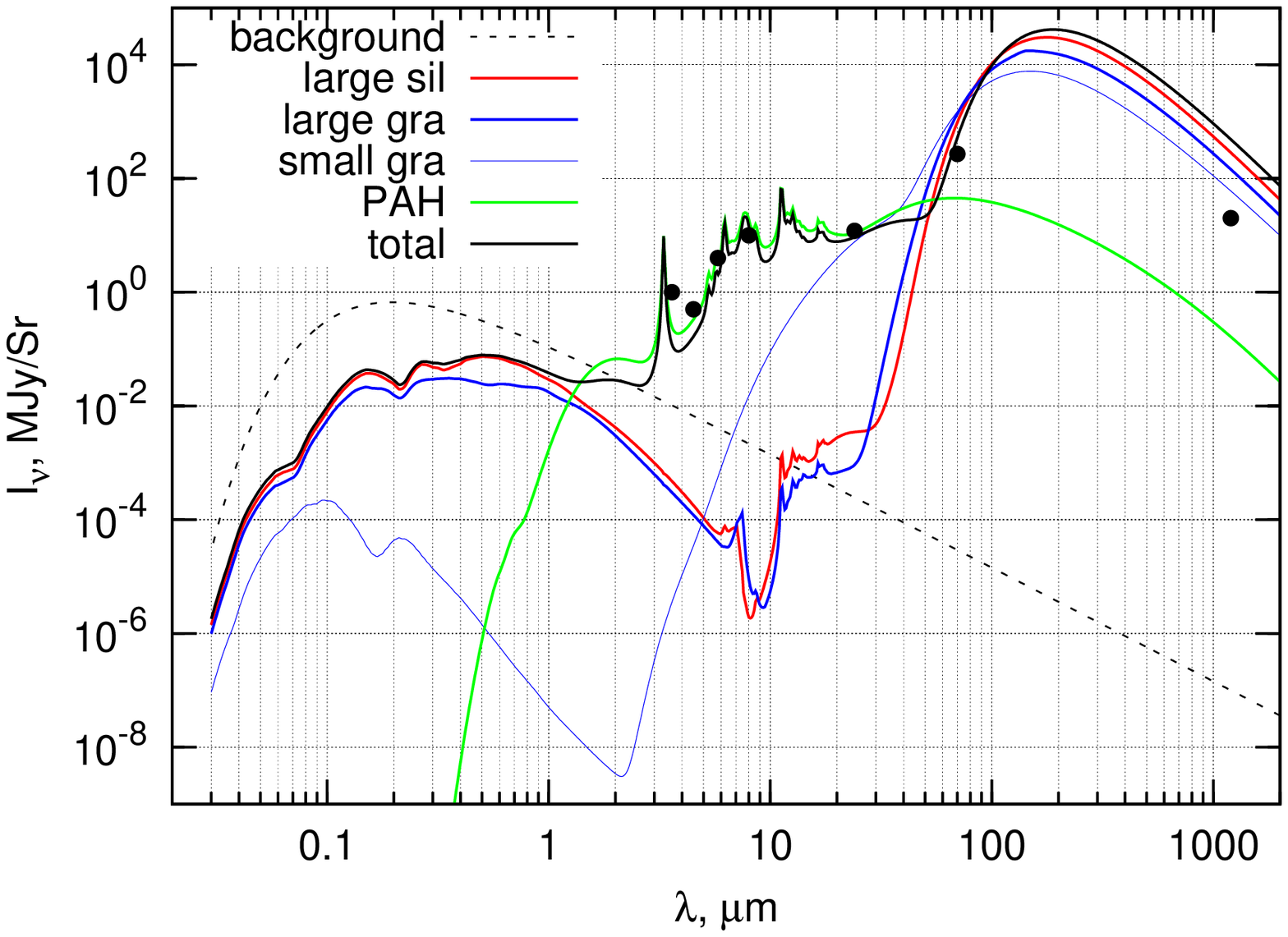,width=0.55\hdsize}}
\caption{Results of RT simulations for the model of IRDC 321 with enhanced interstellar radiation field.}
\label{Tfuncenh}
\end{figure*}

Apparently, stochastic heating by the standard diffuse UV field does not make a
noticeable contribution to the mid-IR spectrum of an IRDC core. The contribution
can be made more significant by increasing the ambient UV irradiation. In Figure
\ref{Tfuncenh} we show same plots as in Figure~\ref{Tfuncnorm}, but computed
with the enhanced ambient UV field, having $T_{\rm bg}=26000$\,K and dilution of
$2\times 10^{-15}$. The relative fraction of hot PAH particles is much higher in
this case, and, as a consequence, the IR radiation intensity toward the core
projected centre (again with 1$^{\prime\prime}$ offset) also increases, this
time matching all the observed values. Computed intensities for the millimetre
emission are higher than observed ones because we do not convolve them with the
beam size that is quite large at this wavelength.

\begin{figure}
\centerline{\psfig{figure=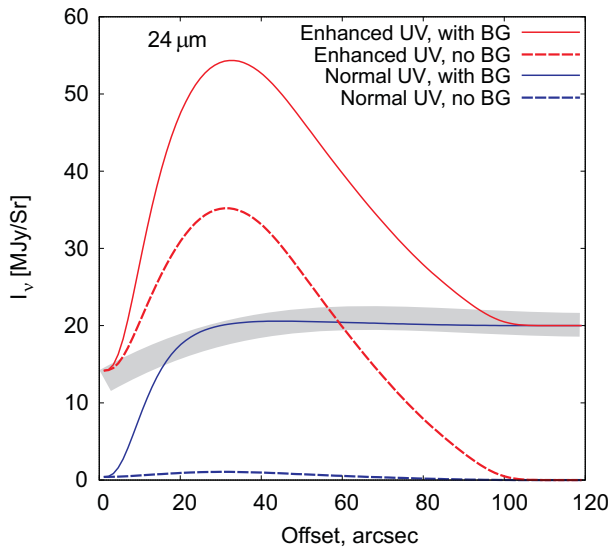,width=0.45\hdsize}}
\caption{Intensity distributions at 24\,$\mu$m over the model core for cases
with and without the background radiation as well as with and without UV
stochastic heating. Observed intensity for the IRDC 321 core is
schematically shown with gray band { (DZF case)}.}
\label{dist}
\end{figure}

\begin{figure}
\centerline{\psfig{figure=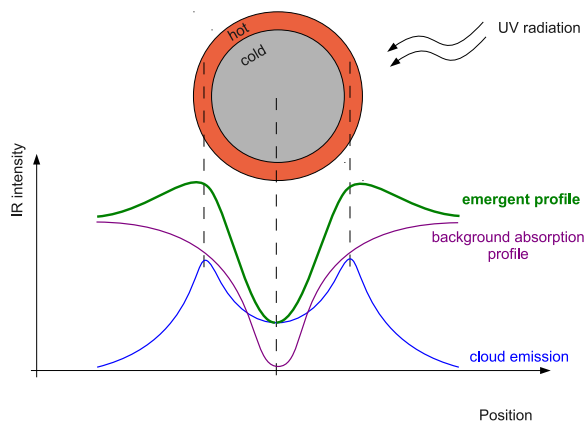,width=0.5\hdsize}}
\caption{Scheme explaining the formation of the ring-like intensity profile in
mid-IR for the core heated by interstellar radiation.}
\label{scheme}
\end{figure}

However, having solved the problem for the central spectrum, we have
simultaneously created an even more severe problem for offset positions. This is
illustrated in Figure~\ref{dist}, where we show distributions of 24\,$\mu$m
intensity for the model core with the standard (blue lines) and enhanced (red
lines) ambient UV fields. With solid lines we show the `true' emergent spectrum
that includes both the attenuated background and the core emission. The proper
core contribution is shown separately with dashed lines. Obviously, for the
standard UV field stochastic heating makes negligible contribution to the
overall core emission (blue dashed line), and the computed intensity at the
projected core centre is much smaller than the observed one. As we make the UV
irradiation stronger, the intensity in the central part does grow up to the
observed value, but at the same time it gets even higher at the core periphery.
The radial intensity profile has a central flat region of about
5$^{\prime\prime}$ in size and reaches the maximum value at 35$^{\prime\prime}$.
This ring-like intensity distribution is formed by stochastically heated small
grains in the envelope as schematically shown in Figure~\ref{scheme}. The
thickness of the  heated layer is low, so the projection effect plays a role in
the formation of the intensity radial distribution. In other words, column
density of heated VSGs and PAHs toward the centre of the core is lower than
toward the core periphery, which results in the intensity difference. The
intensity distribution at 70\,$\mu$m over the core surface is similar to the one
for 24\,$\mu$m except for the strong emission peak toward the location of the
protostar. This peak appears due to lower optical depth at 70\,$\mu$m.

Such ring-like structures, apparently, are not ubiquitous in IRDCs, so we must
admit that stochastic heating do not seem to be a viable explanation for the
alleged excess mid-IR emission in IRDC cores. But the problem would be solved if
we can find some other dust heating mechanism which would work not only in the
core envelope but in its whole volume. Among various non-radiative dust heating
mechanisms, there is one that is often included in chemical models of prestellar
cores. Specifically, these models include mantle evaporation due to stochastic
heating of dust grains by cosmic ray particles \citep{leger}. In the following
subsection, we check if same dust heating that is responsible for mantle
destruction can change appreciably the core SED.

\subsection{Stochastic heating by cosmic rays}

The interaction of cosmic rays (CR) with interstellar medium is a complex
process. The collision of highly energetic CR particles with interstellar
molecules results in formation of secondary particles (pions, mesons, positrons,
gamma-rays, etc.) that also interact with interstellar gas. Both primary and
secondary non-thermal particles dissociate and ionize interstellar molecules as
well as collide with dust grains and contribute to their stochastic heating.
Ideally, to calculate stochastic heating of dust grains induced by cosmic rays
one should know 1) the flux/energy distribution of all non-thermal particles,
and 2) the fraction of the particle kinetic energy that goes into the grain
thermal energy upon the collision. The proper modelling of these processes is a
challenging problem that is further restricted by uncertainties of involved
physical data.

Here we use a simplified approach. The most significant contribution to
stochastic heating of dust grains is assumed to come from free electrons formed
due to propagation of cosmic rays. The mean energy of free electrons produced by
cosmic ray ionization is 20-35\,eV \citep{ev20}. We assume that this energy
entirely goes to the thermal energy of a dust grain in a single collision. We
adopt the standard ionization rate $10^{-17}$\,s$^{-1}$ and assume that these
electrons do not interact with gas and only collide with dust grains. Since the
last assumption is unrealistic (the total effective cross-section of molecules
is much higher than the total effective cross-section of dust grains), we
significantly overestimate the effect.

\begin{figure*}
\centerline{\psfig{figure=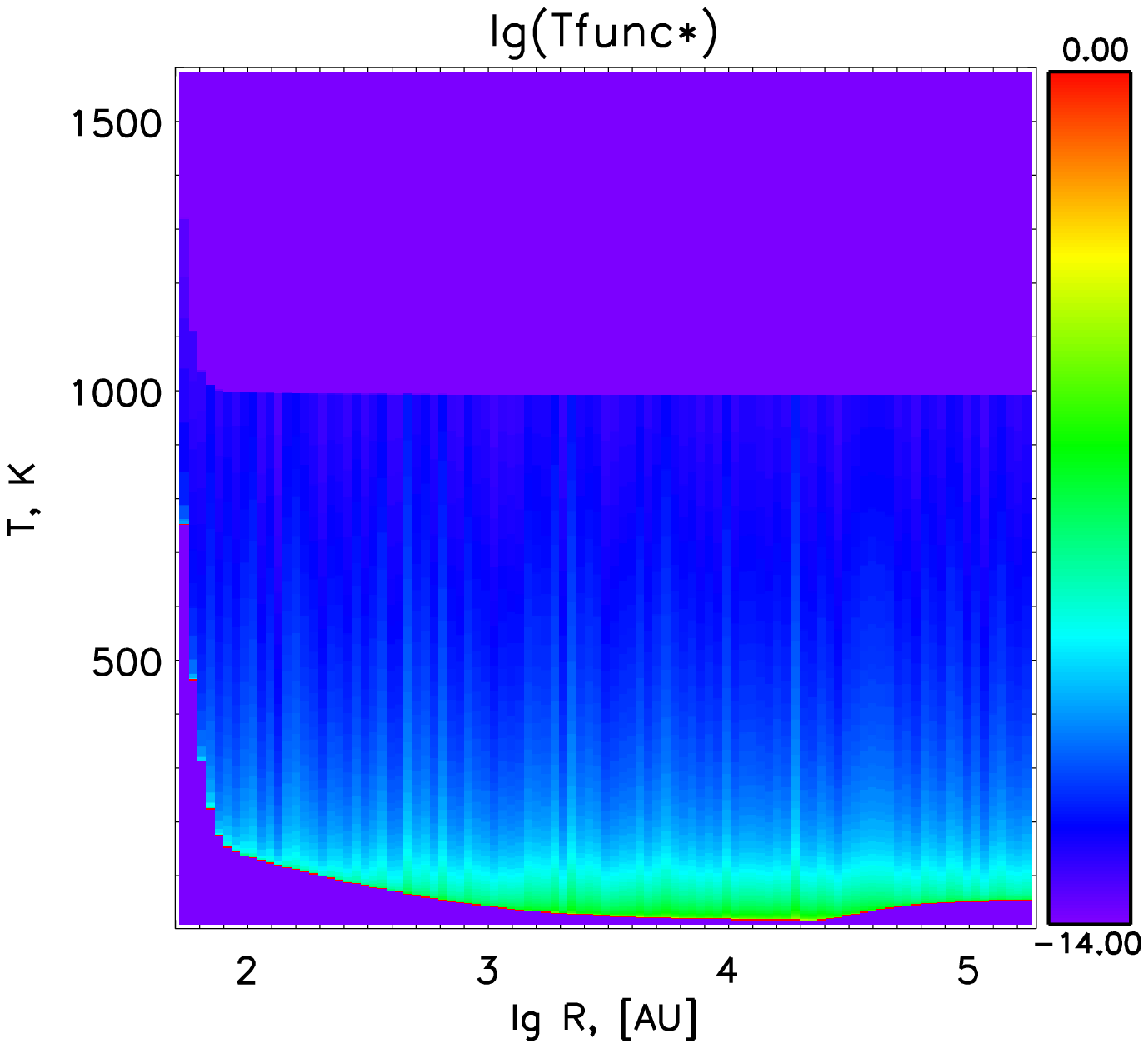,width=0.45\hdsize}
\psfig{figure=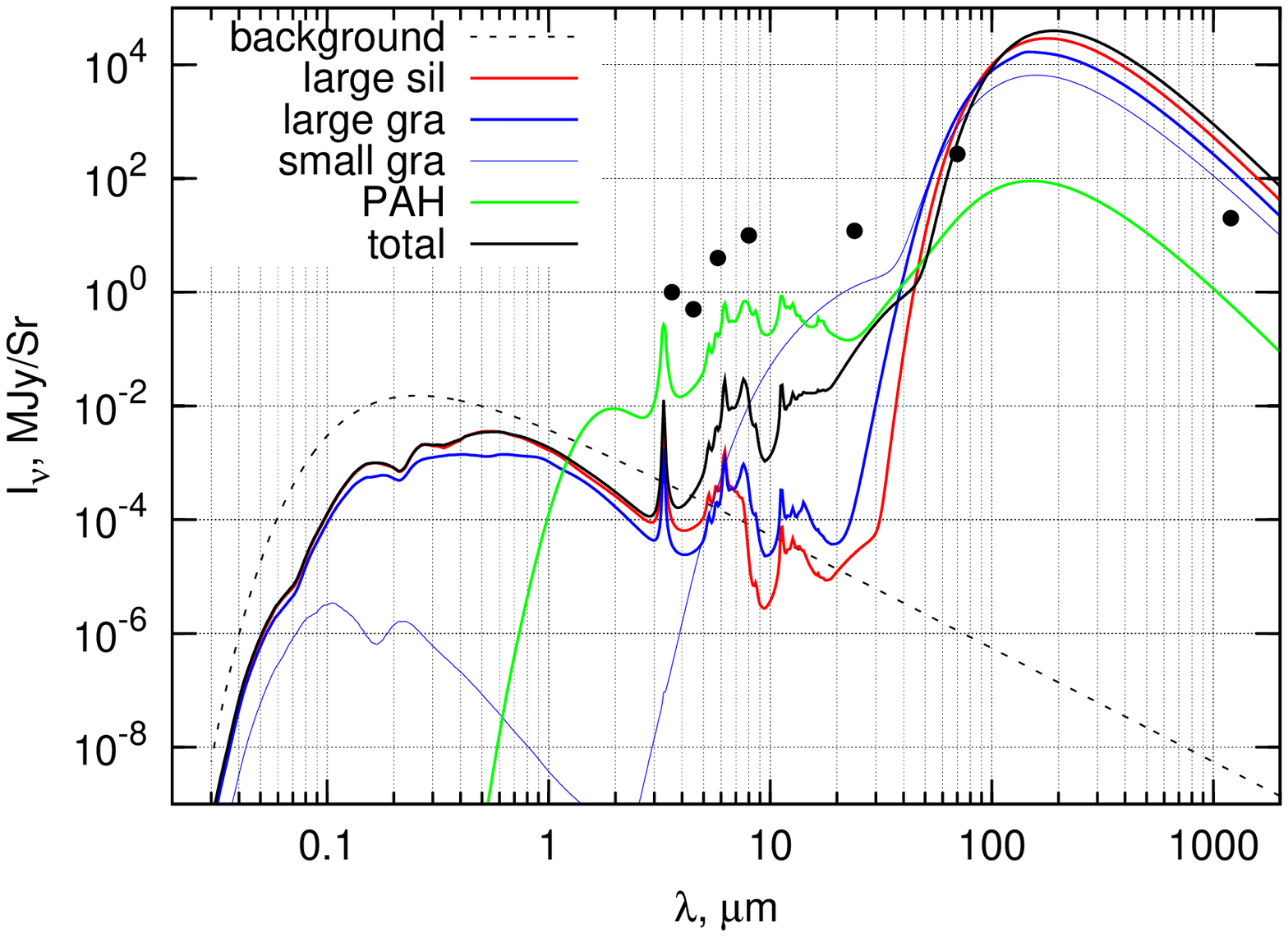,width=0.55\hdsize}}
\caption{Same as in Figure \ref{Tfuncnorm}, but for the model with CR stochastic heating.}
\label{SEDcr}
\end{figure*}

In Figure~\ref{SEDcr} we present $P(T)$ distribution for PAH particles in the
model with CR stochastic heating. To isolate the CR effect, we treat heating by
external UV radiation in the integral non-stochastic way which provides the
minimum temperature for grains. While PAHs are sometimes heated up to 1000\,K,
the probability to find a particle with such a temperature is low. In other
words, heating events are very rare, and PAHs spend most of the time at the
lowest temperature. The contribution of grains, stochastically heated by CR, to
the SED is even smaller than in the case of standard UV (right panel of
Figure~\ref{SEDcr}). The computed 24\,$\mu$m intensity is only a few hundredth
MJy/ster, while the observed intensity is about twenty MJy/ster. Given the
significant overestimation of the collision rate between electrons and grains in
our model, we conclude that stochastic heating of dust grains due to CR cannot
solve the problem of high mid-IR intensity toward IRDC cores.

\section{Discussion}

The analysis of IRDC mid-IR properties indicates that these clouds may possess
intrinsic emission in this range. On a larger scale, it is typically assumed
that emission in the 24\,$\mu$m Spitzer band is produced by stochastically
heated small grains. Our study shows, however, that on a smaller scale
stochastic dust heating by UV photons and CR particles cannot solve the problem
of excess 24\,$\mu$m emission, that \cite{Pavlyuchenkov2011} have inferred for
some IRDC cores. 

For a standard interstellar UV field the contribution of stochastically heated
grains to the core mid-IR emission is negligible. If we try to overcome this
problem by enhancing the ambient UV field, the intensity distribution becomes
ring-like. { The reason behind this is quite simple and qualitatively similar
to the ``limb brightening" effect considered by \cite{Leung1989} for longer
wavelengths (140--300\,$\mu$m) which are not affected by stochastic heating
of small grains.} The UV irradiation of the
core creates the overheated dust layer on its surface. Geometrically, this layer
is narrow due to strong UV absorption. On the other hand, it is nearly
transparent in the mid-IR range so that the hot surface of the dense core should
be observed as a bright rim. Such structures are not seen in any of the cores we
have checked in this study. While 24\,$\mu$m rings are actually observed in
other sources, most of them are classified as planetary nebulae and
circumstellar shells around massive (proto)stars
\citep{Mizunoetal2010,Wachteretal2010}. We conclude that the stochastic heating
of small particles, at least, in its typical implementation, does not seem to be
a viable candidate for the source of mid-IR emission in IRDC cores. Other known
potential stochastic heating factor, cosmic rays, is also not effective, even if
somewhat extreme parameters are adopted.

Our inability to fit mid-IR observations may at least in part be related to the
deficiencies of the model. For example, we assume spherical symmetry for our
cores, with smooth density distributions, resulting in large UV optical depths
in the core outer parts. The problem with 24\,$\mu$m emission would probably be
alleviated if UV radiation is able to penetrate deeper into the core due to its
irregular structure. In this case instead of the narrow hot layer we would have
a more extended hot envelope. Also, we consider only UV and CR heating, but
another bulk emission generation mechanisms are possible. \cite{chemheat}
suggested that surface chemical reactions can be an energy source that excites
infrared emission bands even in grains of `classical' sizes. Some chemically
induced energy deposition in dust grains is also implied by the non-thermal
mantle desorption mechanism suggested by \cite{garrod}. The presence of some
unaccounted energy source indirectly follows from the low sticking probability
(of about 0.3) found by \cite{cb17} in their detailed study of the CB17 core.
They suggested that this low value may be an indication of some desorption
mechanism which is present in these objects and is different from thermal
desorption and cosmic ray induced desorption. Such a desorption mechanism would
also mean an energy deposition into dust grains.

If the core irregular structure is related to turbulence, then turbulent
dissipation may provide some heating. However, it seems likely that in the
absence of strong shocks this mechanism would rather somewhat raise the mean
temperature of grains, enhancing the core emission in the far-IR and
submillimetre bands, but not in the near- and mid-IR.

Yet another factor that is capable to affect the mid-IR core emission is
scattering. In this study, we assume that scattering is isotropic. If our
assumption is relaxed in favour of forward scattering we would have some
additional contribution from the core central hot region. However, the inferred
core structure implies that it is opaque at 24\,$\mu$m, so even with forward
scattering the contribution from the inner region should not be important.

Scattering of the ambient infrared emission can be significant if we assume some
grain evolution in the cores. Extended mid-IR emission in the 3.6 and
4.5\,$\mu$m bands, that was termed MIR cloudshine and coreshine by \cite{jurgen}
and \cite{pagani}, observed in some molecular clouds and cores, was interpreted
as a result of scattering of the ambient infrared light by large grains. In
principle, non-zero $E$ due to scattering at 24\,$\mu$m is also possible. In
this study we consider only four representative grain types and do not vary
their parameters along the core radius. If we would include very large grains
($a>10\,\mu$m) in the model,  scattering at 24\,$\mu$m would be greater, making
mid-IR shadows less prominent.  { However, these large grains would need to
be present in sufficient amount such that scattering dominated over the
absorption by small grains.} It is also possible that scattering on large porous
dust aggregates, like those considered by \cite{ormeletal2011}, may contribute
into observed mid-IR flux.

Obviously, by varying dust optical properties and size distribution along the
core radius, we may achieve a better fit. However, these parameters cannot be
varied in a unconstrained way. Grains in the interstellar medium can grow by
accretion of refractory elements \citep{voshen} or volatile species but the
potential for this growth is limited by the abundance of heavy elements. So, the
preferred way for grains to grow to large sizes is coagulation. In other words,
the number of large grains can only increase at the expense of small grains and
PAH particles. The two-fold outcome of this process is that scattering becomes
more effective, but at the same time contribution of stochastic heating
diminishes. To estimate the physically motivated balance between these two
processes, one needs to include a grain growth in the model. In general, the
search for dust parameters is a complex problem which has to be solved in a
self-consistent way, using observations at as many wavelengths as possible. Such
a modelling should be a subject of a dedicated study.

\section{Conclusions}

Recently, \cite{Pavlyuchenkov2011} found that 24\,$\mu$m emission in mid-IR
shadows of typical IRDC cores does not fit into the combined emission and
absorption picture, being too bright for the density and temperature
distributions that are needed to explain intensity maps of the cores in other IR
and millimetre bands. This problem arises if the only foreground that is
subtracted from the signal is the zodiacal light contribution estimated from the
DIRBE data. The natural way out of this is to assume that the remaining `extra'
emission comes from some other foreground source, like intervening interstellar
dust. However, analysis of observational data shows that the typical foreground
contribution is about { 40\% or less} for nearby objects. All the remaining
emission should originate in the object itself.

We checked if stochastic heating of small grains can produce this extra
24\,$\mu$m emission. However, it only allows to reproduce the central mid-IR
intensity for the studied core if we assume that the core is illuminated by the
diffuse UV light enhanced by a factor of 70 relative to the average Galactic
value. In this case, intensity at the core edges also grows due to projection
effects so that the entire core acquires a ring-like appearance which does not
seem to be observed.  { Stochastic heating by cosmic ray particles does not
change the SED significantly due to the low energy density input of these
particles.}

Overall, we conclude that the origin of at least some mid-IR emission in IRDC
dark cores is unclear and requires a special study. A great care must be taken
when someone uses infrared observations to deduce IRDC properties and to compare
them with results of numerical modelling. The assumption that IRDCs are
completely dark in the infrared (so that all the emission is the foreground
emission) does allow to reproduce the observational data, but in this case {\em
all\/} the information on the proper core emission is lost which may lead to
wrong inferences.

\section*{Acknowledgments}

We are grateful to the referee for constructive suggestions. This study was
supported by the RFBR grant 10-02-00612 and the President of the RF grant
MK-3651.2012.2.  We thank B. Shustov, A. Stutz, R. Launhardt, and C. Dullemond
for useful discussions. The code computing dust optical properties was provided
by D. Semenov (MPIA, Heidelberg, Germany). YP thanks Natalia Kudryavtseva for
stimulating conversations.

\appendix
\section{Details of stochastic heating algorithm}

Here we provide details of the method that is used to calculate the
temperature evolution of an isolated dust grain exposed to
the radiation field with the mean intensity $J_{\nu}$. Emission
(radiative cooling) of the dust grain is described as a
continuous process, while absorption (radiative heating) is represented
either as a discrete process or as a continuous process,
depending on the photon energy. The concept of continuous
cooling is a reasonable approximation to the problem \citep{Draine2001}.

First, we split the spectrum of the external radiation field
into the low-energy and high-energy intervals
separated by a critical frequency $\nu_{\rm c}$, defined by the relation
$h\nu_{\rm c}=0.01E_{\rm th}$, where $E_{\rm th}$ is the mean thermal energy of the
dust grain calculated under the assumption of continuous
heating and cooling.  We use stochastic treatment
only for photons from the high-energy interval of the
spectrum since low-energy photons does not
produce significant temperature fluctuations of the grain.

The next step is to evaluate the time and
frequency sequence of the absorbed photons within the
interval $(\nu_{\rm c}, \nu_{\rm max})$, where $\nu_{\rm max}=10^{16}$\,Hz is the
adopted maximum frequency of photons.
We consider $M$ absorption events and generate the frequency
sequence $\{\nu_1,...,\nu_M\}$ using the Monte Carlo
simulation for the re-normalized probability density distribution of the
{\em absorbed\/} photons
\begin{equation}
p(\nu) = \dfrac{Q_{\nu}^{\rm abs}J_{\nu}}{h\nu},
\end{equation}
where $Q_{\nu}^{\rm abs}$ is the absorption efficiency factor for a given
grain type. The corresponding time sequence $\{t_1,...,
t_M\}$ of the absorption events is simulated using the
Poisson statistics
\begin{equation}
f(t) = \lambda \exp (-\lambda t),
\end{equation}
where $f(t)$ is the probability density distribution for the time interval $t$ between
successive events, $\lambda = M/t_{0}$ is the mean number of
events per unit time, $t_0$ is the total length of the
sequence. The total time $t_0$ can be determined from the
relation
\begin{equation}
\pi a^2 t_{0} \int\limits_{\nu_{c}}^{\nu_{\rm max}}{Q_{\nu}^{\rm abs}J_{\nu}\,d\nu} = 
\dfrac{1}{4\pi}\sum_{i=1}^{M} h\nu_{i},
\end{equation}
which represents the energy absorbed by the grain.
The set of obtained sequences $\{\nu_i\}$ and $\{t_i\}$
is the discrete representation of the energy deposited into the grain during
time $t_0$ by photons from ($\nu_{\rm c}, \nu_{\rm max}$).

Next we calculate the
temperature jump of the grain due to the photon absorption. Let us
suppose that the temperature of the grain just before the
absorption is $T_0$. The temperature $T_1$ just
after the absorption of a photon with energy $h\nu$ is given by
\begin{equation}
U(T_1) - U(T_{0}) = h\nu,
\label{fig_abs}
\end{equation}
where $U(T)$ is the thermal energy of the grain with temperature $T$.
The relation between thermal energy and temperature
is non-linear since the heat capacity of the grain, $C_{\rm V}$,
is a function of temperature. We approximated heat capacities,
presented in Figure~2 of \cite{Draine2001}, by the simple
phenomenological law
\begin{equation}
\dfrac{C_{\rm V}}{Nk} = \dfrac{3}{1+\left(\dfrac{T_{\rm d}}{T}\right)^2},
\end{equation}
where $N$ is the number of atoms in the grain, $k$ is Boltzmann constant,
the heat capacity parameter $T_{\rm d}$ is 175\,K for silicate and 450\,K for graphite. The corresponding
thermal energy as a function of temperature
\begin{equation}
\dfrac{U}{Nk} = 3\left(T-T_{\rm d} \arctan{\frac{T}{T_{\rm d}}} \right).
\label{A6}
\end{equation}
{ Equation \eqref{A6} is substituted into equation \eqref{fig_abs}, which is then
solved for $T_1$ using the bisection method.}

\begin{figure}
\centerline{\epsfig{figure=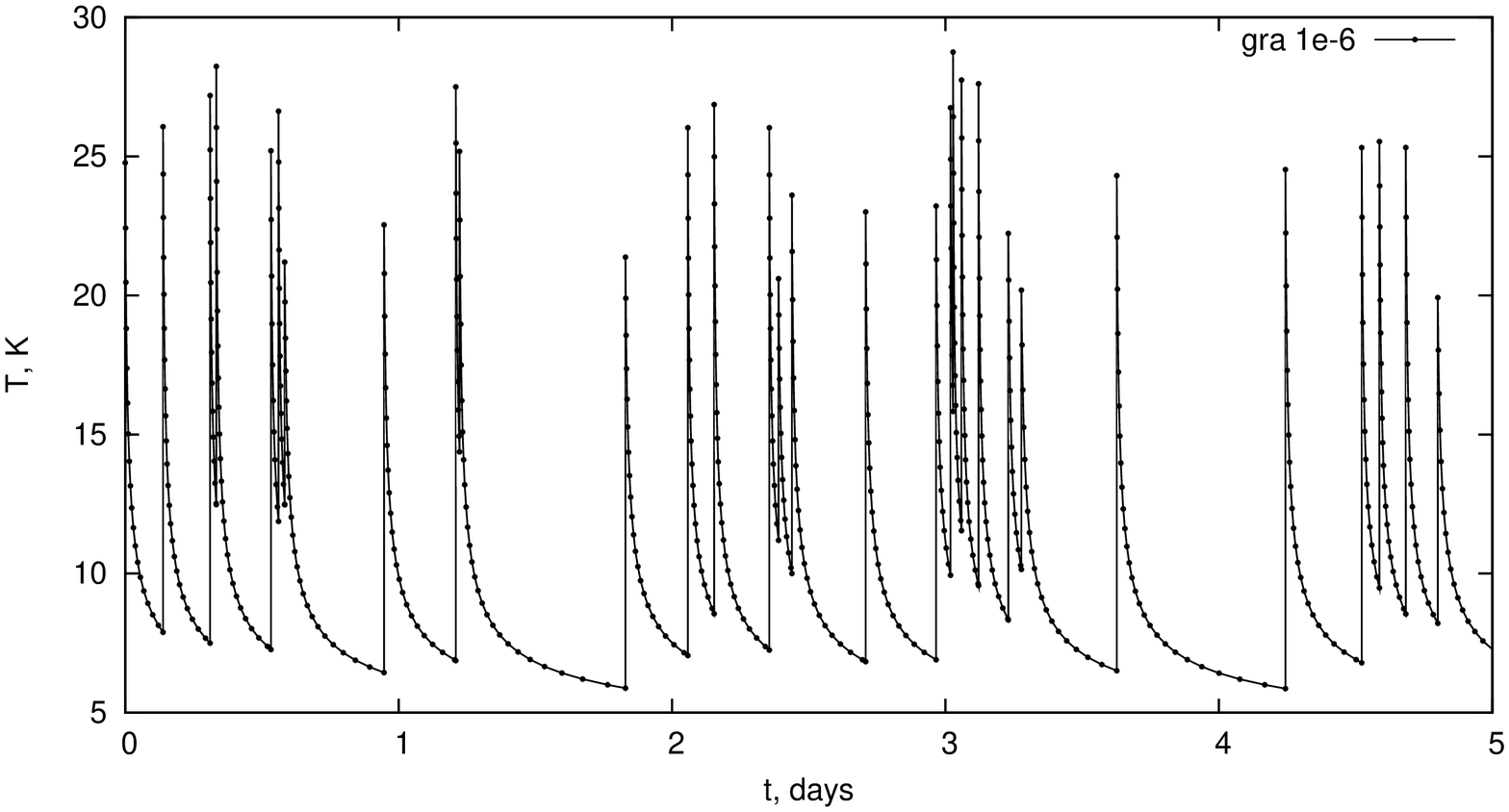,width=0.45\hdsize}}
\caption{
The temperature evolution of a $10^{-6}$\,cm graphite grain
exposed to the Planck radiation with $T_{\rm bg}=20000$\,K and dilution of $10^{-16}$.}
\label{fig_excurs}
\end{figure}

In order to calculate the temperature evolution of the dust grain
between two subsequent absorption events we solve the following equation
\citep[e.g.,][]{Kruegel2003}:
\begin{equation}
\dfrac{dU(T)}{dt} = 4\pi\pi a^2
\left(\int\limits_0^{\nu_{\rm c}}Q_{\nu}^{\rm abs}J_{\nu}d\nu - 
\int\limits_0^{\nu_{\rm max}}Q_{\nu}^{\rm abs}B_{\nu}(T)d\nu \right).
\end{equation}
The first term on the right hand side is the energy absorbed from
the radiation field in the low-energy frequency interval per unit time,
while the second term is the cooling rate due to continuous
radiation. We solve this equation using an implicit Euler
method where the corresponding finite-difference equation
is solved by the bisection method. Since the solution
of this equation describes a fast temperature decay right after the
absorption followed by the slow temperature evolution, we use
adaptive time step control which allows having
about 30 time-steps to evaluate the temperature evolution
between absorptions. An example of the temperature history
for a small graphite grain is shown in Figure~\ref{fig_excurs}.

\begin{figure}
\centerline{\epsfig{figure=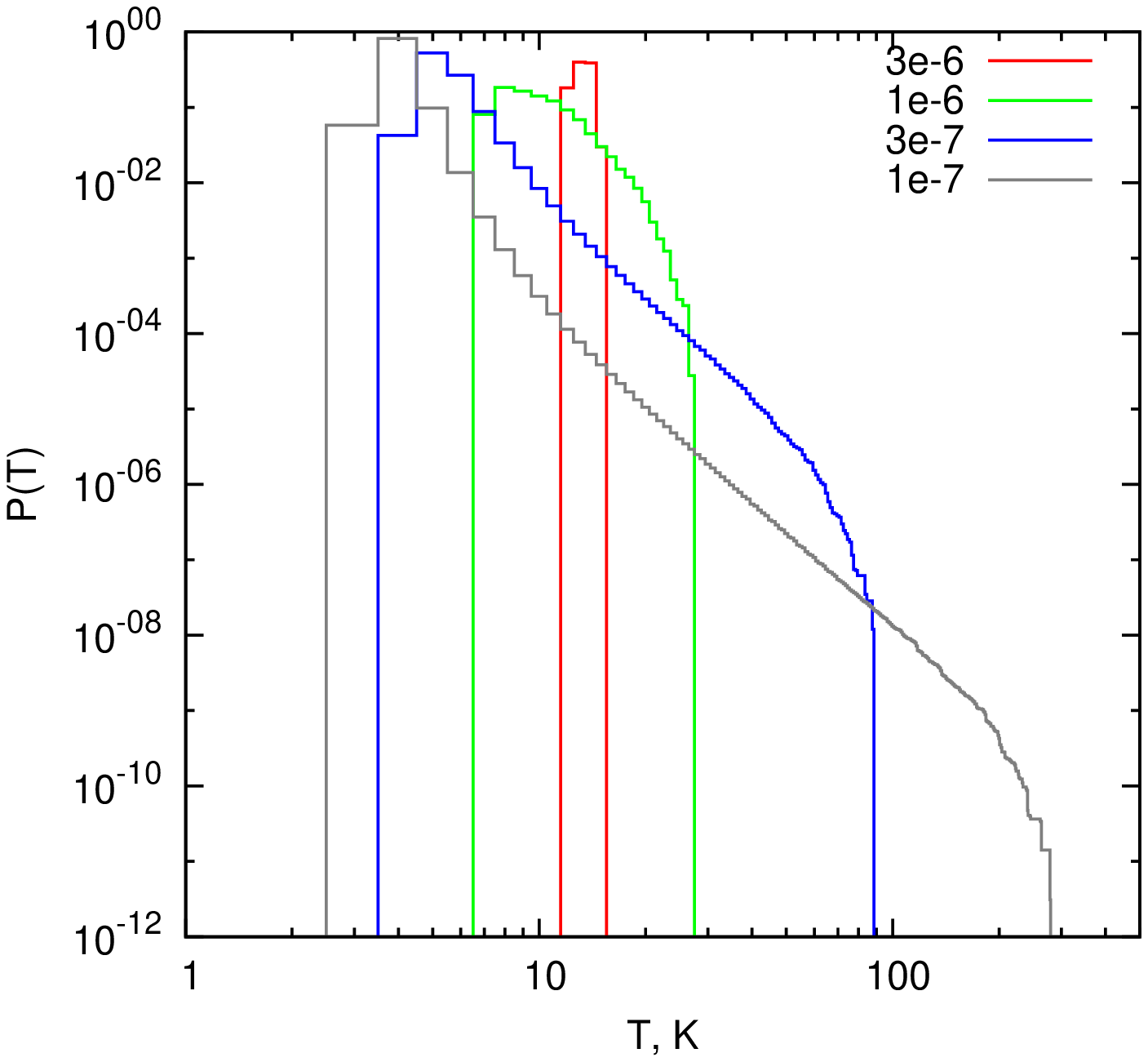,width=0.45\hdsize}}
\caption{The temperature probability density distribution for graphite grains
of different radii exposed to the Planck radiation
with $T_{\rm bg}=20000$\,K and dilution of $10^{-16}$.}
\label{fig_P}
\end{figure}

The last step of the algorithm is to convert the history
$T(t)$ into the temperature probability density distribution $P(T)$ assuming the ergodic hypothesis. We split
the temperature into the number of intervals and calculate the relative
time that the grain spends in each interval. An example
of $P(T)$ for different grain sizes is shown in Figure~\ref{fig_P}.

\label{lastpage}
\end{document}